\documentclass[%
 reprint,
 amsmath,amssymb,
 aps,
pra,
prb,
]{revtex4-1}

\usepackage{subfigure}
\usepackage{graphicx}
\usepackage{dcolumn}
\usepackage{bm}
\usepackage{hyperref}

\begin{document}

\preprint{APS/123-QED}

\title{Generalized Optical Signal Processing Based on Multi-Operator Metasurfaces Synthesized by Susceptibility Tensors   }

\author{Ali Momeni, Hamid Rajabalipanah, Ali Abdolali}
\author{Karim Achouri  }%





\begin{abstract}
This paper theoretically proposes a multichannel functional metasurface computer characterized by Generalized Sheet Transition Conditions (GSTCs) and surface susceptibility tensors.  The study explores a polarization- and angle-multiplexed metasurfaces enabling multiple and independent parallel analog spatial computations when illuminated by differently polarized incident beams from different directions. The proposed synthesis overcomes substantial restrictions imposed by previous designs such as large architectures arising from the need of additional subblocks, slow responses, and most importantly, supporting only the even reflection symmetry operations for normal incidences, working for a certain incident angle or polarization, and executing only single mathematical operation. The versatility of the design is demonstrated in a way that an ultra-compact, integrable and planar metasurface-assisted platform can execute a variety of optical signal processing operations such as spatial differentiation and integration. It is demonstrated that a metasurface featuring non-reciprocal property can be thought of as a new paradigm to break the even symmetry of reflection and perform both even- and odd-symmetry mathematical operations at normal incidences. Numerical simulations also illustrate different aspects of multichannel edge detection scheme through projecting multiple images on the metasurface from different directions. Such appealing findings not only circumvent the major potential drawbacks of previous designs but also may offer an efficient, easy-to-fabricate, and flexible approach in wave-based signal processing, edge detection, image contrast enhancement, hidden object detection, and equation solving without any Fourier lens. 
\end{abstract}

\pacs{Valid PACS appear here}
\maketitle


\section{INTRODUCTION}
The idea of analog computing was pioneered at the early 19th century, where traditional analog computers or calculating machines executed general mathematical operations, mechanically or electronically \cite{goodman2005introduction,hausner1971analog,maclennan2007review,stark2012application}. Nevertheless, these solutions have actually suffered from the disadvantages of slow response and relatively large size so that they were then quite secluded by evolving digital computing schemes in the rapidly changing technology age \cite{nakamura2016image,solli2015analog}. Regarding the recent flourishing achievements in the apparently unrelated field of metamaterials \cite{wegener2013metamaterials}, analog computations have had a glorious return to the competition as a new concept, because they can escape from some inherent restrictions of their digital counterparts such as data conversion loss \cite{solli2015analog}. Motivated by the recent renewed interest in metamaterial-based analog computing ( Silva, A. et.al., Science 2014, 343, 160-163) \cite{silva2014performing}, the new emerging analog computing can be basically split into two major categories of temporal and spatial computation. The proposals based on the time domain calculations, usually lead to remarkably large devices that are very difficult to integrate \cite{yang2014all,li2009arbitrary,slavik2008photonic,tan2013high}. In the spatial domain, such computing enables massively parallel processing of entire images with no energy cost, which provides significant advantages against standard digital processing of images \cite{pors2014analog}. To go beyond the drawbacks of temporal analog computing, two metasurface and Green's Function (GF) approaches have emerged for performing metamaterial-based spatial analog computations \cite{babashah2017integration,pors2014analog,abdollahramezani2017dielectric}. In the first method in which the operator of choice is implemented by a thin planar metamaterial in Fourier domain, the preliminary necessity of two additional sub-blocks to apply Fourier and Inverse Fourier Transforms, implies a significant increase in the overall size of system. Furthermore, the bulky lenses and filters hinders their practical worth, especially in compact architectures \cite{babashah2017integration,silva2014performing}. By not doing the computations in Fourier domain, this limitation can be circumvented in the second approach pioneered by Silva et al, wherein the GF of a transversally homogeneous but longitudinally inhomogeneous block itself emulates the operator of choice \cite{silva2014performing}. All proposals reported in this way, however, still have major weaknesses such as thick profiles \cite{doskolovich2014spatial,zangeneh2018analog}, supporting only the even symmetry operations for normal incidences \cite{silva2014performing,youssefi2016analog,zhu2017plasmonic,zangeneh2017spatial}, working properly for a certain incident angle or polarization \cite{youssefi2016analog,zhu2017plasmonic}, and  executing only single mathematical operation \cite{kwon2018nonlocal,ruan2015spatial,zhu2018generalized}. Considering the spectacular development in the design and analysis of metasurfaces over the past lustrum, they seem to be good candidates for being engineered to implement the spatial transfer functions of different desired mathematical operations \cite{zhu2017plasmonic,pors2014analog,abdollahramezani2015analog}. 

Metasurfaces, the dimensional reductions of volumetric metamaterials and the functional extensions of frequency selective surfaces, are made of an engineered array of  subwavelength particles, enabling the transformation of the incident waves into the desired reflected and transmitted waves \cite{achouri2015general,achouri2016metasurface}. Compared to the three-dimensional (3D) bulky metamaterials, metasurfaces benefits from smaller physical footprint, lower losses, and easier fabrication processes while providing greater flexibility and novel functionalities \cite{chen2016review}. Because of opening new possibilities to control the amplitude, phase, and polarization of the electromagnetic (EM) waves with a very thin profile, they are known as peculiar and flexible hosts for many applications such as tunable/broadband absorptions \cite{rahmanzadeh2018multilayer,mehrabi2018polarization,rahmanzadeh2018adopting}, invisibility cloaking \cite{ni2015ultrathin}, scattering manipulation \cite{momeni2018information,rouhi2018real}, antenna engineering \cite{rahmanzadeh2018adopting}  and so on. At the microscopic level, metasurfaces are efficiently modeled as a complex sheets being typically much thinner than the operating wavelength. They cannot be described by conventional boundary conditions and current commercial software as they may support both electric and magnetic field discontinuities and exhibit arbitrary bianisotropy \cite{achouri2016metasurface}. A few analytical schemes based on equivalent impedance tensors \cite{asadchy2016perfect,epstein2016synthesis,epstein2017unveiling,kwon2018arbitrary}, momentum transformation \cite{salem2014manipulating}, and polarizability of particles \cite{niemi2013synthesis,ra2017metagratings} have been recently explored for metasurface synthesis. Alternatively, to handle full vectorial problems and giving closed-form solutions, Achouri et al. proposed a general and fast metasurface design method based on surface susceptibility tensors and Generalized Sheet Transition Conditions (GSTCs) \cite{achouri2015general,achouri2015synthesis,achouri2016metasurface,lavigne2018susceptibility,achouri2016comparison,achouri2018design}. This strategy, which generally describes the metasurface discontinuity in terms of an expansion over derivatives of the Dirac delta distribution, inherently brings about greater insight into the physics of the problem. Afterwards, relying on such a straightforward synthesis tool, several metasurface concepts and applications were investigated, including EM wave plates \cite{yu2012broadband}, EM surface waves \cite{achouri2018space}, orbital angular momentum multiplexing \cite{achouri2015synthesis}. Although the ability to direct the flow of EM has captured the fascination of scientists and engineers; however, the room is still free for the other noble EM applications like analog computing.\

 \begin{figure}[t]
	\includegraphics[scale=.29]{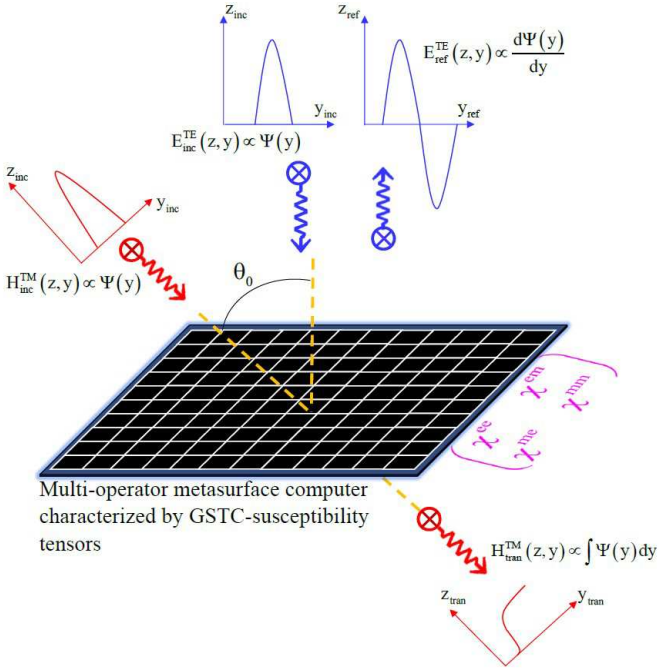}
	\caption{\label{fig:epsart} The proposed multi-operator metasurface spatial computer characterized by GSTC-susceptibility tensors.}
\end{figure}
Motivated by the renewed interest in optical signal processing, we propose and demonstrate a multi-operator metasurface computer whose underling working principle relies on GF method. 
 It is shown that using electric and magnetic surface susceptibilities synthesized by GSTCs, a properly designed versatile metasurface can completely lift the aforementioned restrictions accompanied with the current optical signal processing technologies. It carries out the required computations in spatial domain in either reflection or transmission region, without any additional bulky subblocks. It is demonstrated that the presence of normal susceptibility components and nonreciprocal features is required for performing angle-multiplexed computing and breaking the even reflection symmetry, respectively. In order to clarify the concept, we present several illustrative examples whereby different wave-based mathematical functionalities including polarization-independent, polarization-multiplexed, and angle-multiplexed spatial analog computing as well as breaking the even reflection symmetry for normally incident waves have been theoretically investigated. In addition, it is demonstrated that the proposed multi-operator metasurface computer can elaborately perform multichannel one-dimensional (1D) edge detection upon illumination by multiple independent images coming from different directions. The proposed edge detection scheme does not require the fine-tuning of geometrical and material properties or bulkiness caused by phase-shift Bragg grating \cite{zengerle1995phase} while its parallel computations significantly accelerate the image processing. In summary, multi-operator metasurfaces synthesized by feasible susceptibility tensors allows us to go beyond most restrictions of current metamaterial-based analog computers within much smaller volumes.

\section{ANALOG COMPUTING BASED ON SUSCEPTIBILITY-GSTCS APPROACH: FUNDAMENTAL THEORY }

Our idea for performing spatial analog computations via metasurfaces synthesized by susceptibility-GSTCs approach is graphically sketched in Fig. 1. Here, we assume that the employed metasurface , which consists of a finite array of planar scattering particles located in the $x-y$ plane $(z=0)$, behaves as a 2D EM discontinuity of subwavelength thickness $(\sigma<<\lambda)$. The purpose of the metasurface synthesis  is to derive a proper set of surface susceptibility components $({{\overline{\overline{\chi }}}_{\text{ee}}},{{\overline{\overline{\chi }}}_{\text{em}}},{{\overline{\overline{\chi }}}_{\text{me}}},{{\overline{\overline{\chi }}}_{\text{mm}}})$ which can transform an arbitrary input wave characterized by field distribution $E_{\text{inc}}^{\text{TE/TM}}\left( {{y}_{\text{inc}}},{{z}_{\text{inc}}} \right)=\psi _{\text{inc}}^{\text{TE/TM}}({{y}_{\text{inc}}})\exp (-j{{k}_{z}}{{z}_{\text{inc}}})$, into arbitrary specified reflected or transmitted field-profiles $E_{\text{ref/tran}}^{\text{TE/TM}}({{y}_{\text{ref}},{z}_{\text{ref}}})=\psi _{\text{ref/tran}}^{\text{TE/TM}}({{y}_{\text{ref/tran}}})\exp (j{{k}_{z}}{{z}_{\text{ref/tran}}})$. In this way, $\psi _{\text{inc}}^{\text{TE/TM}}({{y}_{\text{inc}}})$ and $\psi _{\text{ref/tran}}^{\text{TE/TM}}({{y}_{\text{ref/tran}}})$ refer to the electric field for TE- or TM-polarized incident and reflected/transmitted beams, respectively.  The incident beam propagates in the negative $z_{\text{inc}}$-direction {in the incident beam coordinate system $(x_{\text{inc}},y_{\text{inc}})$ and is impinging on the metasurface at an angle $\theta$ with respect to $z$ axis {in the metasurface coordinate system $(x,z)$}. Accordingly, the homogeneous metasurface will reflect/transmit the beam toward the $z_{\text{ref}}/z_{\text{tran}}$-direction [ in the reflected/transmitted beam coordinate system $(x_{\text{ref}},y_{\text{ref}})/(x_\text{tran},y_\text{tran})$]. Through the vectorial spatial Fourier transform, the beam can be expressed through its angular spectrum as the superposition of plane waves with different directions, i.e. \cite{zhu2017plasmonic}. $E_{\text{inc}}^{\text{TE/TM}}\left({{y}_{\text{inc}}},{{z}_{\text{inc}}} \right)=\int{\tilde{\psi }_{\text{inc}}^{\text{TE/TM}}({{k}_{y}})\exp \,(j{{k}_{y}}{{y}_{\text{inc}}}-j{{k}_{z}}{{z}_{\text{inc}}})d{{k}_{y}}}$ 
	where, $\tilde{\psi }_{\text{inc}}^{\text{TE/TM}}({{k}_{y}})$ denotes the angular spectrum of the incident beam in which the tilde indicates the Fourier transform, $\theta $ is the incident angle and ${{k}_{y}}={{k}_{0}}\sin \theta $, and ${{k}_{z}}={{k}_{0}}\sqrt{1-{{\sin }^{2}}\theta }$ are the wavevector components of each plane wave, respectively. Suppose that a Gaussian beam profile with the narrow bandwidth of $W<<k_0 $ in the spatial Fourier domain impinges onto the metasurface discontinuity with the incident wave angle of $\theta_0$ from $z>0$. 
	The transformation of the profile of the beam upon interacting with the metasurface computer can be described in terms of the linear system theory \cite{youssefi2016analog}, i.e. with the spatial spectral transfer function of the operator of choice. Indeed, our approach is based on the fact that the linear convolution $\psi _{\text{ref/tran}}^{\text{TE/TM}}({{y}_{\text{ref}}})=\int{\psi _{\text{inc}}^{\text{TE/TM}}(u)\,g_{\text{ref/tran}}^{\text{TE/TM}}({{y}_{\text{ref}}}-u)du}$ between an arbitrary incident field distribution, $\psi _{\text{inc}}^{\text{TE/TM}}(y)$, and the GF associated with the operator of choice, $g_{\text{ref/tran}}^{\text{TE/TM}}(y)$, may be expressed in the spatial Fourier space as $\overline{\overline{{\tilde{{G}}}}}_{\text{ref/trans}}^{\text{TE/TM}}\left( {{k}_{y}} \right)\,\tilde{\psi }_{\text{inc}}^{\text{TE/TM}}({{k}_{y}})$.  The transfer functions describing the transformation of the beam profile in reflection and transmission have the form of $\tilde{\hat{{H}}}_{\text{ref/tran}}^{\text{TE/TM}}\left( {{k}_{y,\text{inc}}} \right)=\overline{\overline{{\tilde{G}}}}\,_{\text{ref/tran}}^{\text{TE/TM}}\left( {{k}_{y}}\left( {{k}_{y,\text{inc}}} \right) \right)$, where
\begin{equation}
\overline{\overline{{\tilde{G}}}}_{\text{ref}}^{\text{TE/TM}}({{k}_{y}})\equiv \left[ \begin{matrix}
{{{\tilde{R}}}^{\text{TE/TE}}}\left( {{k}_{y}} \right)\, & {{{\tilde{R}}}^{\text{TE/TM}}}\left( {{k}_{y}} \right)\,\,  \\
{{{\tilde{R}}}^{\text{TM/TE}}}\left( {{k}_{y}} \right)\,\, & {{{\tilde{R}}}^{\text{TM/TM}}}\left( {{k}_{y}} \right)\,  \\
\end{matrix} \right]\,\,\,
\end{equation}

\begin{equation}
\overline{\overline{{\tilde{G}}}}_{\text{tran}}^{\text{TE/TM}}({{k}_{y}})\equiv \left[ \begin{matrix}
{{{\tilde{T}}}^\text{{TE/TE}}}\left( {{k}_{y}} \right)\, & {{{\tilde{T}}}^{\text{TE/TM}}}\left( {{k}_{y}} \right)\,\,  \\
{{{\tilde{T}}}^{\text{TM/TE}}}\left( {{k}_{y}} \right)\,\, & {{{\tilde{T}}}^{\text{TM/TM}}}\left( {{k}_{y}} \right)\,  \\
\end{matrix} \right]\,\,\,
\end{equation}

and ${{k}_{y}}\left( {{k}_{y,\text{inc}}} \right)={{k}_{0}}\sin \left( \theta +{{\theta }_{0}} \right)\approx {{k}_{y,\text{inc}}}\cos {{\theta }_{0}}+{{k}_{0}}\sin {{\theta }_{0}}$ by assuming $\left| \theta  \right|\le si{{n}^{-1}}({W}/{{{k}_{0}}})$ \cite{youssefi2016analog}. 	 $\tilde{R}\left( {{k}_{y}}\left( {{k}_{y,\text{inc}}} \right) \right)\,,\,\,\tilde{T}\left( {{k}_{y}}\left( {{k}_{y,\text{inc}}} \right) \right)$ also represent the anisotropic complex reflection and transmission coefficients of the meta-atoms constituting of the metasurface computer for the angle of incidence $\theta +{{\theta }_{0}}$ . Therefore,    the reflection/transmission coefficients calculated for various incident angles, can be mapped onto a transfer function in the spatial Fourier domain.   According to the above relations, the profiles of the transmitted and reflected beams in respective coordinate systems can be written as  \cite{youssefi2016analog,zhu2017plasmonic}.
\begin{align}
& E_{\text{ref}}^{\text{TE/TM}}\left( {{y}_{\text{ref}}},{{z}_{\text{ref}}} \right)=\\\nonumber&\int{\tilde{\hat{H}}_{\text{ref}}^{\text{TE/TM}}\left( {{k}_{y,\text{inc}}} \right)\psi _{\text{inc}}^{\text{TE/TM}}\left( {{k}_{y}}\left( {{k}_{y,\text{inc}}} \right) \right)} \\ \nonumber
& exp \,(j{{k}_{y}}{{y}_{\text{ref}}}+j{{k}_{y}}{{z}_{\text{ref}}})\,d{{k}_{y,\text{inc}}} \\ \nonumber 
\end{align}
\begin{align}
& E_{\text{tran}}^{\text{TE/TM}}\left( {{y}_{\text{tran}}},{{z}_{\text{tran}}} \right)=\\\nonumber&\int{\tilde{\hat{H}}_{\text{tran}}^{\text{TE/TM}}\left( {{k}_{y,\text{inc}}} \right) \psi _{\text{inc}}^{\text{TE/TM}}\left( {{k}_{y}}\left( {{k}_{y,\text{inc}}} \right) \right)} \\\nonumber &exp \,(j{{k}_{y}}{{y}_{\text{tran}}}+j{{k}_{y}}{{z}_{\text{tran}}})\,d{{k}_{y,\text{inc}}} \\\nonumber 
\end{align}

The GSTCs formulation treats the metasurface in terms of electric (P) and magnetic (M) surface polarization densities, where the susceptibility tensors are related to the incident, reflected and transmitted fields around the surface. In the absence of impressed sources, the GSTCs read \cite{achouri2015general} 

\begin{align}
& \hat{z}\times \Delta \textbf{H}=j\omega {{\textbf{P}}_{\parallel }}-\hat{z}\times {{\nabla }_{\parallel }}{{\textbf{M}}_{z}} \\ 
& \Delta \textbf{E}\times \hat{z}=j\omega \mu {{\textbf{M}}_{\parallel }}-{{\nabla }_{\parallel }}(\frac{{{\textbf{P}}_{z}}}{\varepsilon_0 })\times \hat{z} \\  
& \hat{z}.\Delta \textbf{D}=-\nabla .{{\textbf{P}}_{\parallel }} \\  
& \hat{z}.\Delta \textbf{B}=-\mu_0 \nabla .{{\textbf{M}}_{\parallel }}, \\ \nonumber
\end{align}
in which, the terms on the left-hand sides of the above equations are the differences of the fields on the two sides of the metasurface computer which may be defined as
\begin{equation}
\Delta {{{\psi }}_{u}}=\psi _{u}^{\text{tran}}-\psi _{u}^{\text{inc}}-\psi _{u}^{\text{ref}}\,\,\,\,\,\,\,\,\,\,\,\,\,\,\,\,\,\,\,\,\,\,\,\,\,\,\,\,\,\,\,\,\,\,\,\,\,\,\,\,u=x,y,z
\end{equation} 
Here, $\psi$ refers to any of the fields H, E, D or B. By taking  the coupling between the adjacent scattering particles to account, P and M can be macroscopically characterized by the associated susceptibility tensors in the following form \cite{achouri2015synthesis}

\begin{align}
& \textbf{P}=\varepsilon {{{\bar{\bar{\chi }}}}_{\text{ee}}}{{\textbf{E}}_{av}}+{{{\bar{\bar{\chi }}}}_{\text{ee}}}\sqrt{\mu \varepsilon }{{\textbf{H}}_{av}}, \\ 
& \textbf{M}={{{\bar{\bar{\chi }}}}_{\text{mm}}}{{\textbf{H}}_{av}}+{{{\bar{\bar{\chi }}}}_{\text{ee}}}\sqrt{\frac{\varepsilon }{\mu }}{{\textbf{E}}_{av}}, \\ \nonumber
\end{align}

where, the average fields are defined as 
\begin{align}
{{\psi }_{\text{avg}}}=\frac{\psi _{u}^{\text{tran}}+\psi _{u}^{\text{inc}}+\psi _{u}^{\text{ref}}\,\,\,\,\,\,\,}{2}\,\,\,\,\,\,\,\,\,\,\,\,\,\,\,\,\,\,\,\,\,\,\,\,\,\,\,\,\,\,\,\,\,u=x,y,z
\end{align}
Consequently, the design problem reduces to finding the susceptibilities of a polarizable zero-thickness metasurface implementing our desired tonsorial Green's function. The number of independent unknown susceptibilities can be reduced by enforcing the Lorentz reciprocity conditions if this is in accordance with the design specifications \cite{lindell1994electromagnetic,kong1986electromagnetic}
\begin{align}
\bar{\bar{\chi }}_{\text{ee}}^{T}={{\bar{\bar{\chi }}}_\text{{ee}}},\,\,\,\bar{\bar{\chi }}_{\text{mm}}^{T}={{\bar{\bar{\chi }}}_{\text{mm}}},\,\,\,\bar{\bar{\chi }}_{\text{me}}^{T}=-{{\bar{\bar{\chi }}}_{\text{em}}}
\end{align}

Given that the solution (the reflected and transmitted fields) contains waves of both TE and TM polarizations, under the assumption of no cross-polarization conversion, the surface susceptibility tensors corresponding to the TM-polarized solution can be determined by 
\begin{align}
& \Delta {{\textbf{H}}_{x}}=-j\omega {{\varepsilon }_{0}}(\chi _{\text{ee}}^{yy}{{\textbf{E}}_{y.\text{av}}}+\chi _{\text{ee}}^{yz}{{\textbf{E}}_{z.\text{av}}})-j{{k}_{0}}\chi _{\text{em}}^{xy}{{\textbf{H}}_{y.\text{av}}}, \\ \nonumber
& \Delta {{\textbf{E}}_{y}}=-j\omega {{\mu }_{0}}\chi _{\text{mm}}^{xx}{{\textbf{H}}_{x.\text{av}}}+j{{k}_{0}}(\chi _{\text{em}}^{yx}{{\textbf{E}}_{y.\text{av}}},\chi _{\text{em}}^{zx}{{\textbf{E}}_{z.\text{av}}})\\\nonumber&-\chi _{\text{ee}}^{yz}{{\partial }_{y}}{{\textbf{E}}_{x.\text{av}}}-\chi _{\text{ee}}^{zz}{{\partial }_{y}}{{\textbf{E}}_{z.\text{av}}}-{{\eta }_{0}}\chi _{\text{em}}^{zx}{{\partial }_{y}}{{\textbf{H}}_{x.\text{av}}}. \\
\end{align}
The dual solution for TE-polarized problem can be simply achieved with proper substitutions. Besides, one can conclude that the transformation of the incident beam profile into the reflected/transmitted beam profile by the metasurface computer is described by multiplication by $\overline{\overline{{\tilde{G}}}}_{\text{ref/tran}}^{\text{TE/TM}}({{k}_{y}})$ in the Fourier space. The Taylor series expansion of $\overline{\overline{{\tilde{G}}}}_{\text{ref/tran}}^{\text{TE/TM}}({{k}_{y}})$ yields 

\begin{align}
& \overline{\overline{\tilde{G}}}_{\text{ref/tran}}^{\text{TE/TM}}\left( {{k}_{y}}\left( {{k}_{y,\text{inc}}} \right) \right)=\overline{\overline{{\tilde{G}}}}_{\text{ref/tran}}^{\text{TE/TM}}({{k}_{0}}\sin ({{\theta }_{0}}))+\\\nonumber &{{\left. \frac{\partial \overline{\overline{{\tilde{G}}}}_{\text{ref/tran}}^{\text{TE/TM}}({{k}_{y}})}{\partial {{{\tilde{k}}}_{y}}} \right|}_{{{k}_{0}}\sin ({{\theta }_{0}})}}\times {{k}_{y}}\cos ({{\theta }_{0}})+O(k_{y}^{2}) \\\nonumber 
\end{align}

With assuming zero reflection or zero transmission, $\overline{\overline{{\tilde{G}}}}_{\text{ref/tran}}^{\text{TE/TM}}({{k}_{0}}\sin ({{\theta }_{0}}))=0$, the equation above reduces to $\rho\frac{{{k}_{y}}}{{{k}_{0}}}+O(k_{y}^{2})$ \cite{youssefi2016analog} wherein $\rho$ is a constant determined by the metasurface susceptibilities. In this case, for sufficiently small spatial bandwidth of input signal, the corresponding Green’s function can be estimated well by a linear approximation, i.e. $\overline{\overline{{\tilde{G}}}}_{\text{ref/tran}}^{\text{TE/TM}}({{k}_{y}})\propto {{k}_{y}}$. Consequently, the reflected or transmitted beam profile will be nothing except the first-order derivative of the input signal multiplied by a fixed scale factor. It means that to realize the Green’s function of the first-order differentiator, a zero reflection/transmission at the specified incident wave angle, ${{\theta }_{0}}$, is required with 180 degree phase difference around ${{\theta }_{x}}$. Similarly, the first-order integrator Green’s function, $\overline{\overline{{\tilde{G}}}}_{\text{ref/tran}}^{\text{TE/TM}}({{k}_{y}})\propto {1}/{{{k}_{y}}}\;$, possess infinite amplitude at the specified incident wave angle (${{\theta }_{0}}$) with 180 degree phase difference around it. Nevertheless, the realistic transfer functions are usually limited to one at the origin, rather than being infinite as it should be for an ideal integrator. This detrimental property, which stems from the passivity constraint, prevents the integrator from retrieving the zero harmonic (or the DC component) of the input signal. By the way, our design still allows the efficient integration of spatially wide AC signals. The excluded low-frequency components, however, may be finally added to the output signal to yield the sought-after integrated signal. Besides, it is worth mentioning that when the spatial bandwidth of the input field increases, the performance of the integrator or differentiator will deviate from the ideal case. Furthermore, the spatial analog computer designation can be followed as an inverse problem yielding the metasurface surface susceptibilities in terms of $\overline{\overline{{\tilde{G}}}}_{\text{ref/tran}}^{\text{TE/TM}}({{k}_{y}})$ corresponding to the desired mathematical operations at the transmission and reflection regimes. Subsequently, the continuous design specifications may be finally discretized into subwavelength pixels and realized by suitably designed polarizable meta-atoms (see Fig. 1).              
\section{ILLUSTRATIVE EXAMPLES }
The diversity of possible beam profile transformations using the proposed metasurface synthesis method is virtually infinite. This section investigates a few illustrative examples of practical interest, in order of increasing complexity. Depending on whether we choose the transmitted or reflected field as the output mathematical wave, the general form of integration and differentiation operators necessitate the corresponding Green's function of the metasurface to have a pole and a zero with a local phase change of 180 around the specified incident angle, respectively. For implementing integrators, due to the limited-amplitude transfer function of the passive metasurfaces, the associated Green' function is truncated to be unity with the width of d in the vicinity of the origin to avoid gain requirement, 
\[\overline{\overline{{\tilde{G}}}}_{\text{ref/tran}}^{\text{TE/TM}}({{k}_{y}})= \left\{ \begin{array}{ll}
1 & \mbox{$|t|<d $};\\
{{(j\,{{k}_{y}})}^{-1}} & \mbox{$|t|>d $}.\end{array} \right. \] 
By analyzing the general form of the transfer function in each example, we will deduce the susceptibility conditions enabling either spatial differentiation or integration of the transverse profile of the incident beam. 

\subsection{Single mathematical transformation with polarization-independent specification}
We heuristically start with the most simple and conventional case, i.e. a uniform metasurface only transforming the phase and amplitude of the reflected/transmitted waves. Regarding the existing orthogonality, the pairs of (${{E}^{\text{TE}}},{{H}^{\text{TE}}}$) and 
(${{E}^{\text{TM}}},{{H}^{\text{TM}}}$) are proportional to their incident counterparts and orthogonal duals, thus, the problem can split into TE- and TM-polarized incidences. The problem of single-operation spatial analog computing between specified incident and reflected/ transmitted fields only requires a mono-anisotropic, diagonal, and hence non-gyrotropic and reciprocal metasurface as 
\begin{align}
&\;\;\;\;\;	\overline{\overline{{{\chi }}}}_{\text{ee}}=\left( \begin{matrix}
	\begin{matrix}
	\begin{matrix}
	\chi _{\text{ee}}^{xx}  \\
	0  \\
	0  \\
	\end{matrix} & \begin{matrix}
	0  \\
	\chi _{\text{ee}}^{yy}  \\
	0  \\
	\end{matrix}  \\
	\end{matrix} & \begin{matrix}
	0  \\
	0  \\
	\chi _{\text{ee}}^{zz}  \\
	\end{matrix}  \\
	\end{matrix} \right)\,\\
&	\overline{\overline{{{\chi }}}}_{\text{mm}}=\left( \begin{matrix}
	\begin{matrix}
	\begin{matrix}\chi _{\text{mm}}^{xx}  \\
	0  \\
	0  \\
	\end{matrix} & \begin{matrix}
	0  \\
	\chi _{\text{mm}}^{yy}  \\
	0  \\
	\end{matrix}  \\
	\end{matrix} & \begin{matrix}
	0  \\
	0  \\
	\chi _{\text{mm}}^{zz}  \\
	\end{matrix}  \\
	\end{matrix} \right)
\end{align}
The spatial transformation is not described by using fields, but rather by using the associated tonsorial Green's function. For such a single-operator metasurface, the relations between the components of tonsorial Green's functions and susceptibilities can be written as \cite{achouri2015general}

\begin{align}
\tilde{G}_{\text{tran}}^{\text{TM}}({{k}_{y}})=\frac{{{k}_{z}}(4+\chi _{\text{ee}}^{yy}\chi _{\text{mm}}^{xx}k_{0}^{2}+\chi _{\text{ee}}^{yy}\chi _{\text{mm}}^{zz}k_{y}^{2})}{(2j-\chi _{\text{ee}}^{yy}{{k}_{z}})(\chi _{\text{mm}}^{xx}k_{0}^{2}+\chi _{\text{ee}}^{zz}k_{y}^{2}-2j{{k}_{z}})},\\
\tilde{G}_{\text{tran}}^{\text{TE}}({{k}_{y}})=\frac{{{k}_{z}}(4+\chi _{\text{mm}}^{yy}\chi _{\text{ee}}^{xx}k_{0}^{2}+\chi _{\text{mm}}^{yy}\chi _{\text{mm}}^{zz}k_{y}^{2})}{(2j-\chi _{\text{mm}}^{yy}{{k}_{z}})(\chi _{\text{ee}}^{xx}k_{0}^{2}+\chi _\text{{mm}}^{zz}k_{y}^{2}-2j{{k}_{z}})},\\
\tilde{G}_{\text{ref}}^{\text{TM}}({{k}_{y}})=\frac{2j(\chi _{\text{mm}}^{xx}k_{0}^{2}+\chi _{\text{ee}}^{zz}k_{x}^{2}-\chi _{\text{ee}}^{yy}k_{z}^{2})}{(2j-\chi _{\text{ee}}^{yy}{{k}_{z}})(\chi _{\text{mm}}^{xx}k_{0}^{2}+\chi _{\text{ee}}^{zz}k_{y}^{2}-2j{{k}_{z}})},\\
\tilde{G}_\text{{ref}}^{\text{TE}}({{k}_{y}})=\frac{2j(\chi _{\text{ee}}^{xx}k_{0}^{2}+\chi _{\text{mm}}^{zz}k_{y}^{2}-\chi _{\text{mm}}^{yy}k_{z}^{2})}{(2j-\chi _{\text{mm}}^{yy}{{k}_{z}})(\chi _{\text{ee}}^{xx}k_{0}^{2}+\chi _{\text{mm}}^{zz}k_{y}^{2}-2j{{k}_{z}})}\\\nonumber
\end{align}
Upon insertion of the spatial derivation condition for both TE and TM polarizations, one can obtain a simple and straightforward but not unique solution as 
\begin{align}
&G_{\text{tran}}^{\text{TM,TE}}=0:\\&\left\{ \begin{matrix}
4+\chi _{\text{ee}}^{yy}\chi _{\text{mm}}^{xx}k_{0}^{2}+\chi _{\text{ee}}^{yy}\chi _{\text{ee}}^{zz}k_{y}^{2}=0\text{   TM polarization}  \\
4+\chi _{\text{mm}}^{yy}\chi _{\text{ee}}^{xx}k_{0}^{2}+\chi _{\text{mm}}^{yy}\chi _{\text{mm}}^{zz}k_{y}^{2}=0\text{ TE polarization}  \nonumber
\end{matrix} \right.
\end{align}
	\begin{figure*}%
	\centering
	\subfigure[]{%
		\label{fig:21}%
		\includegraphics[height=1.7in]{Fig_2a}}%
	\subfigure[]{%
		\label{fig:22}%
		\includegraphics[height=1.6in]{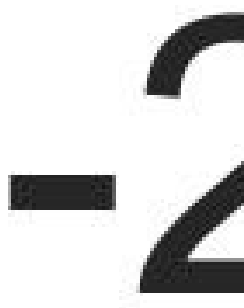}}%
	\qquad
	\subfigure[]{%
		\label{fig:23}%
		\includegraphics[height=2.8in]{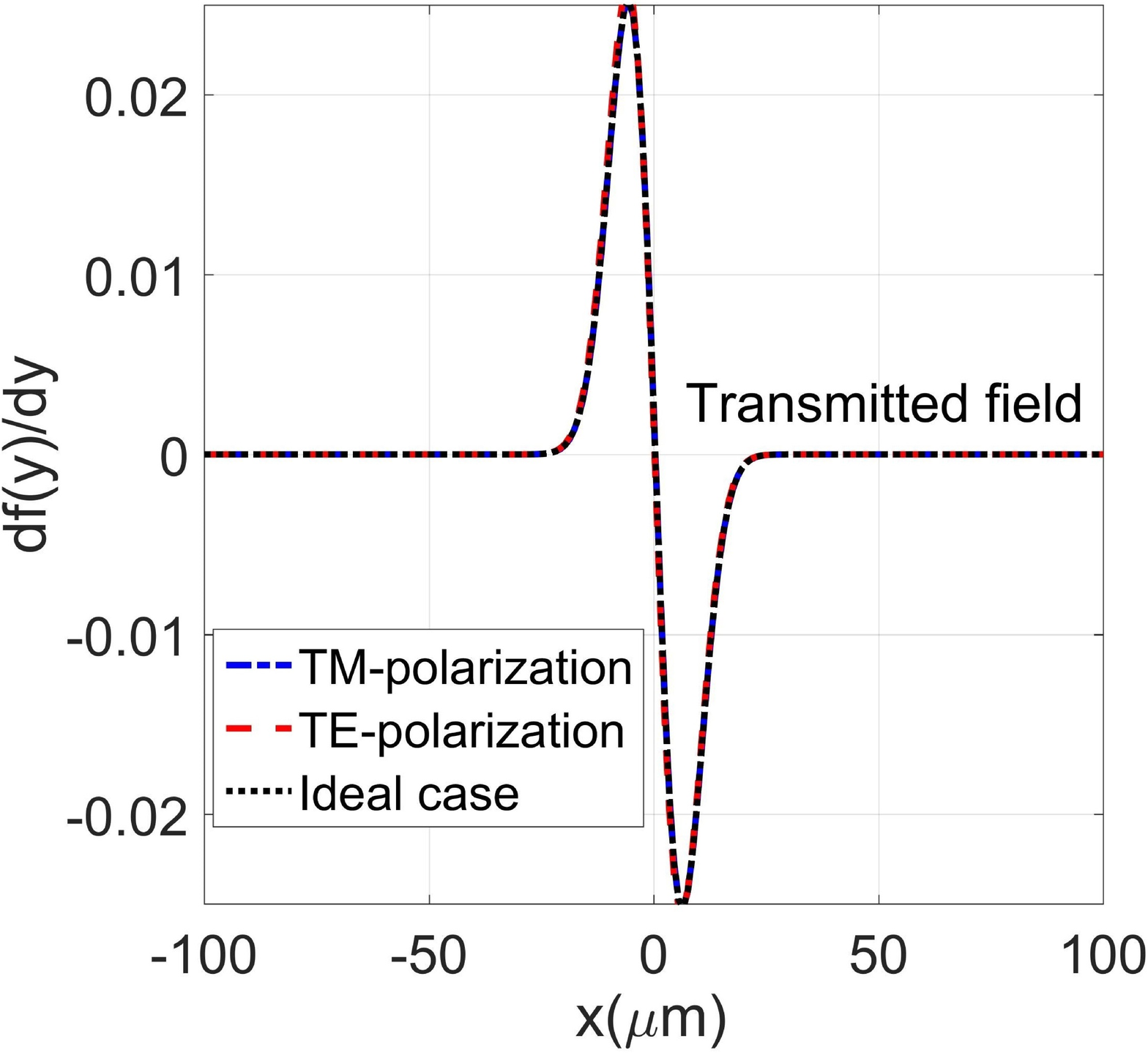}}%
	\subfigure[]{%
		\label{fig:24}%
		\includegraphics[height=2.8in]{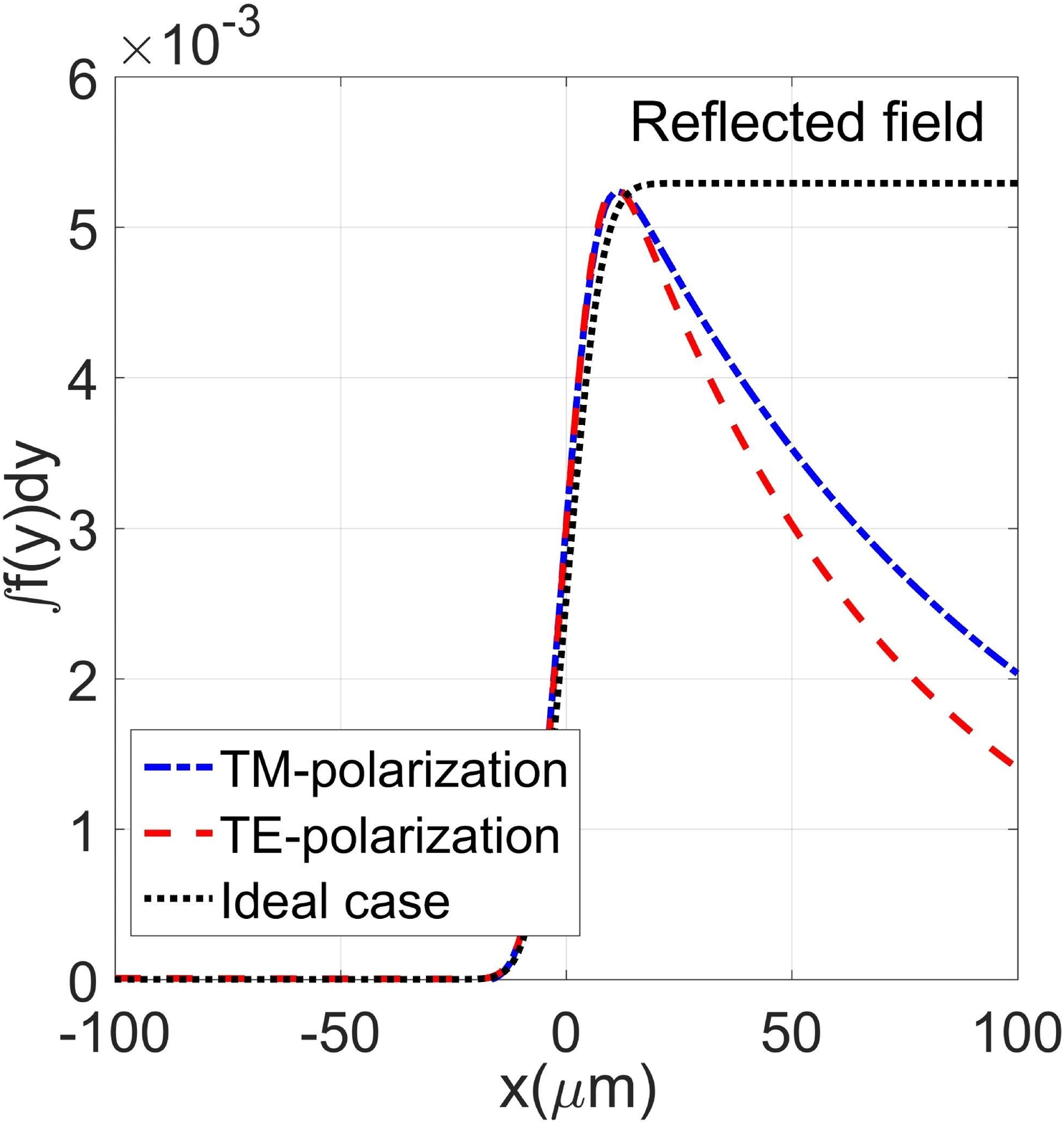}}%
	\caption{ (a) TE-polarized incident field profile together with reflected derivative field profile. (b) TE-polarized incident field profile together with reflected integrated field profile. The reflected field together with scaled calculated first-order (c) derivative and (d) integration of the input field for both TE and TM polarizations.    }
	\label{fig:4}
\end{figure*}
To inspect the dual-polarized performance of spatial derivation and integration functionalities, we consider a Gaussian-shaped electric field with a spatial bandwidth of $W=0.1k_0$ obliquely impinging onto the metasurface computer. Let us now look at the single field-profile transformations of Fig. 2 established to synthesize our single-operator metasurface computer. The first  
transformation illustrated in Fig. 2a, consists of an oblique TM-polarized incident Gaussian beam being differentiated and transmitted at ${{\theta }_{0}}={{45}^{{}^\circ }}$. The high degrees of freedom provided by the GSTCs-susceptibility synthesis method allows us to go beyond the previous restrictions arising from non-zero transmission coefficients of longitudinally finite slabs \cite{youssefi2016analog,zangeneh2018analog}. The second transformation shown in Fig. 2b, also consists in the integration and fully transmission of a TE-polarized incident beam profile impinging on the metasurface at ${{\theta }_{0}}={{45}^{{}^\circ }}$. Fig. 3 depict the exact and the synthesized Green's functions of metasurface related to the derivation and integration operators, respectively, whose trends successfully overlap each other around ${{k}_{y}}=0$. It is obvious that the Green's function synthesized for performing derivation has a first order zero and that of integration has a first order pole at ${{\theta }_{0}}={{45}^{{}^\circ }}$. However, due to the previously explained reason, its associated value is truncated to unity. Numerical results depicted in Fig. 3 verify reasonable accordance between analytical solution of the desired transfer function pertaining to the operators of choice and the transverse electric field distributions of both TE and TM polarizations. The calculated transverse profiles belonging to the spatial derivation and integration of the incident beam as well as those of reflected/transmitted waves are plotted in Figs. 2c, d. These figures show good agreement between the rigorously calculated beam profiles and the beam profiles resulted by the synthesized metasurfaces. For instance, the accuracy of the differentiation is 99.3\%, described by the Pearson correlation coefficient between the simulated and analytical computed reflected/transmitted field amplitudes.\

 As can be seen, while the integrator has worked well to retrieve the non-zero spatial harmonics of the input field, it fails to retrieve the zero harmonic or the DC part. A modified method \cite{babashah2017integration}, can be used to improve the accuracy of integration when the metasurface is fed by signals rich in low-frequency contents. As the inset of Fig. 2 displays, the designed single-operator metasurface functions similarly upon illumination by the incident fields of both TE and TM polarizations. Note that, in this section, we have considered only three different sets of susceptibilities to perform the same transformation. There are other possible combinations of susceptibilities that may be considered to perform this operation. However, most of them correspond to unpractical scenarios of being lossy, active, and/or only perform the specified transformation for one polarization state of the incident wave \cite{achouri2015general}.   
	\begin{figure*}%
	\centering
	\subfigure[]{%
		\label{fig:221}%
		\includegraphics[height=2.5in]{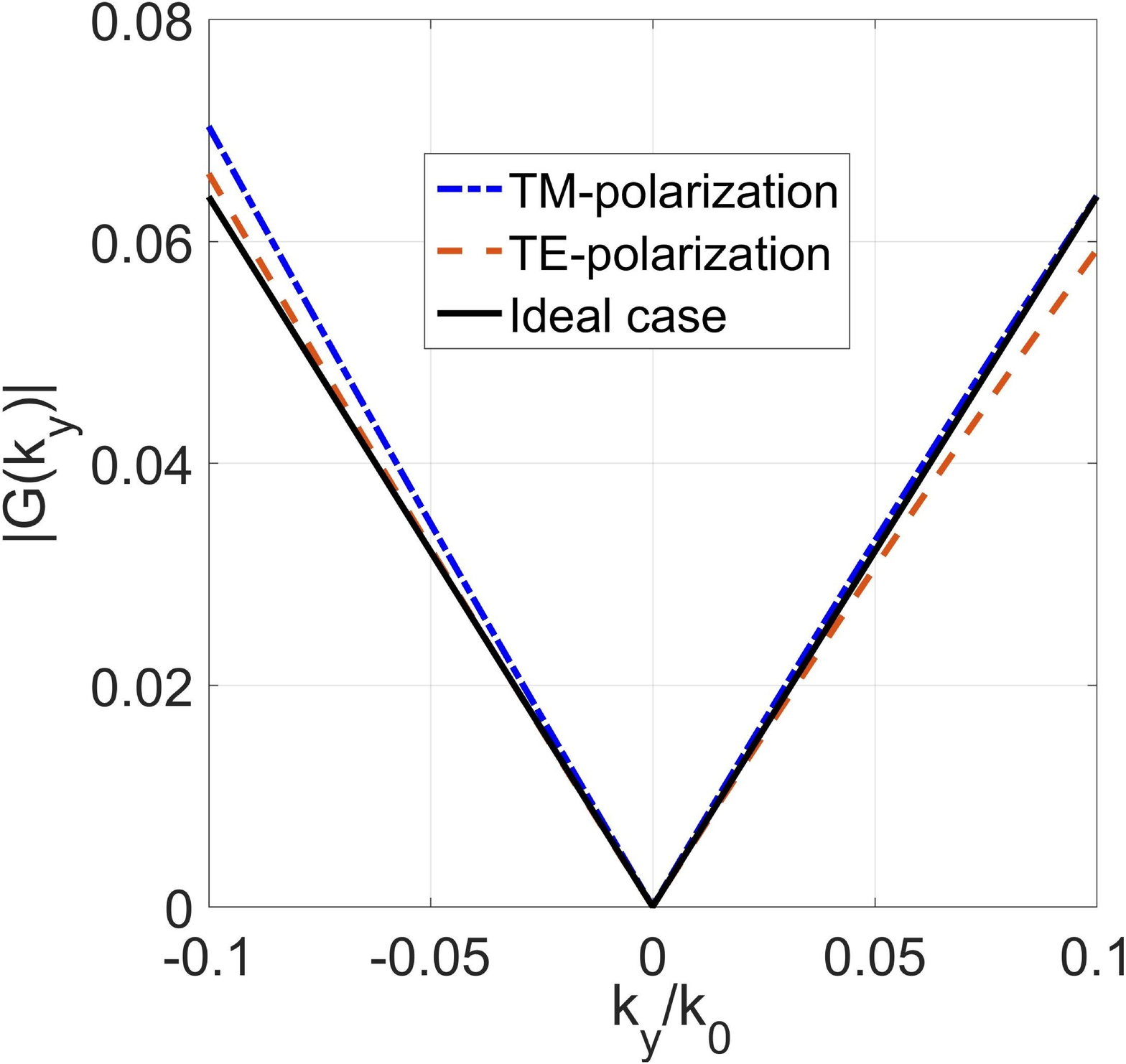}}%
	\subfigure[]{%
		\label{fig:222}%
		\includegraphics[height=2.5in]{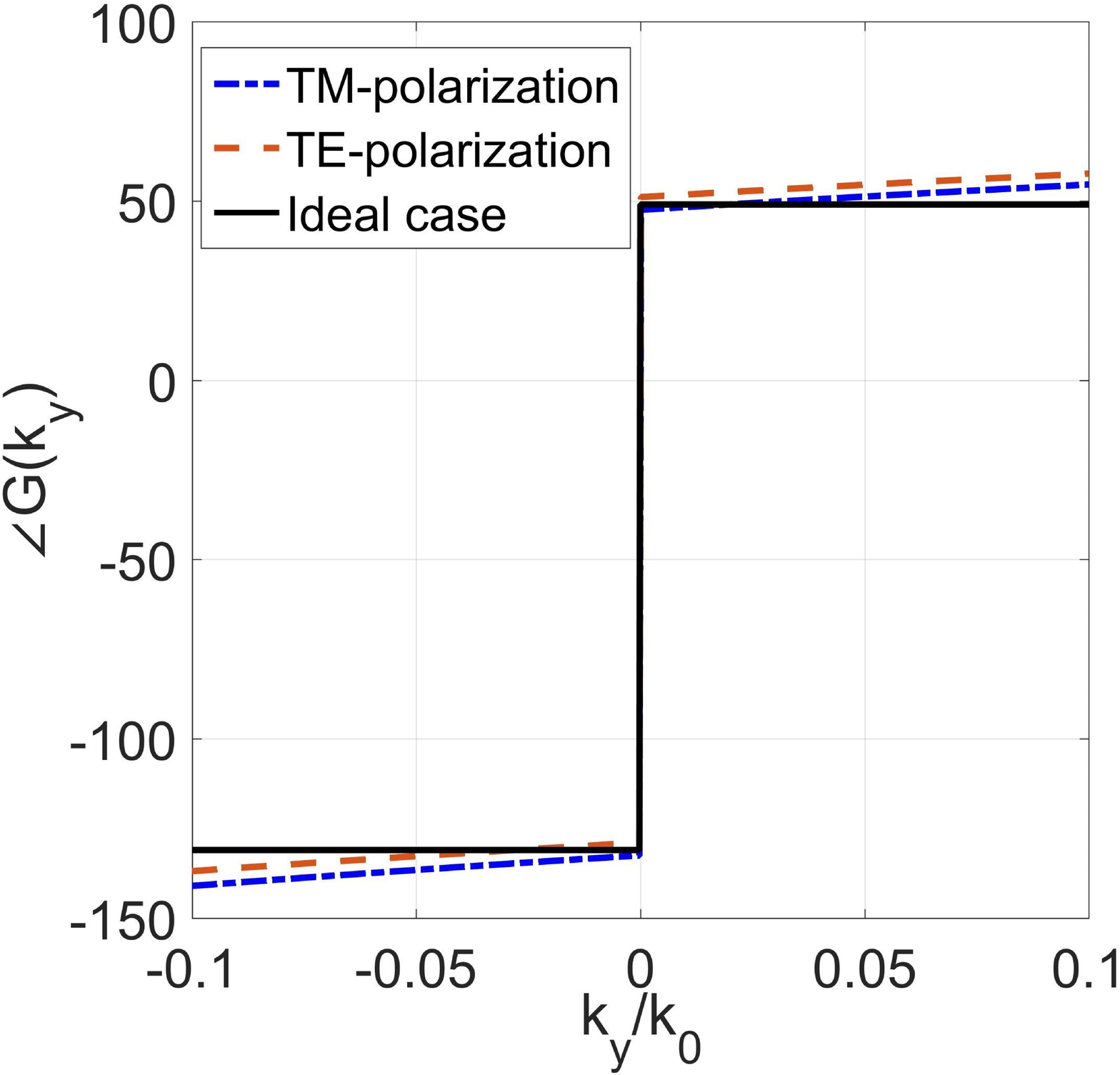}}%
	\qquad
	\subfigure[]{%
		\label{fig:223}%
		\includegraphics[height=2.5in]{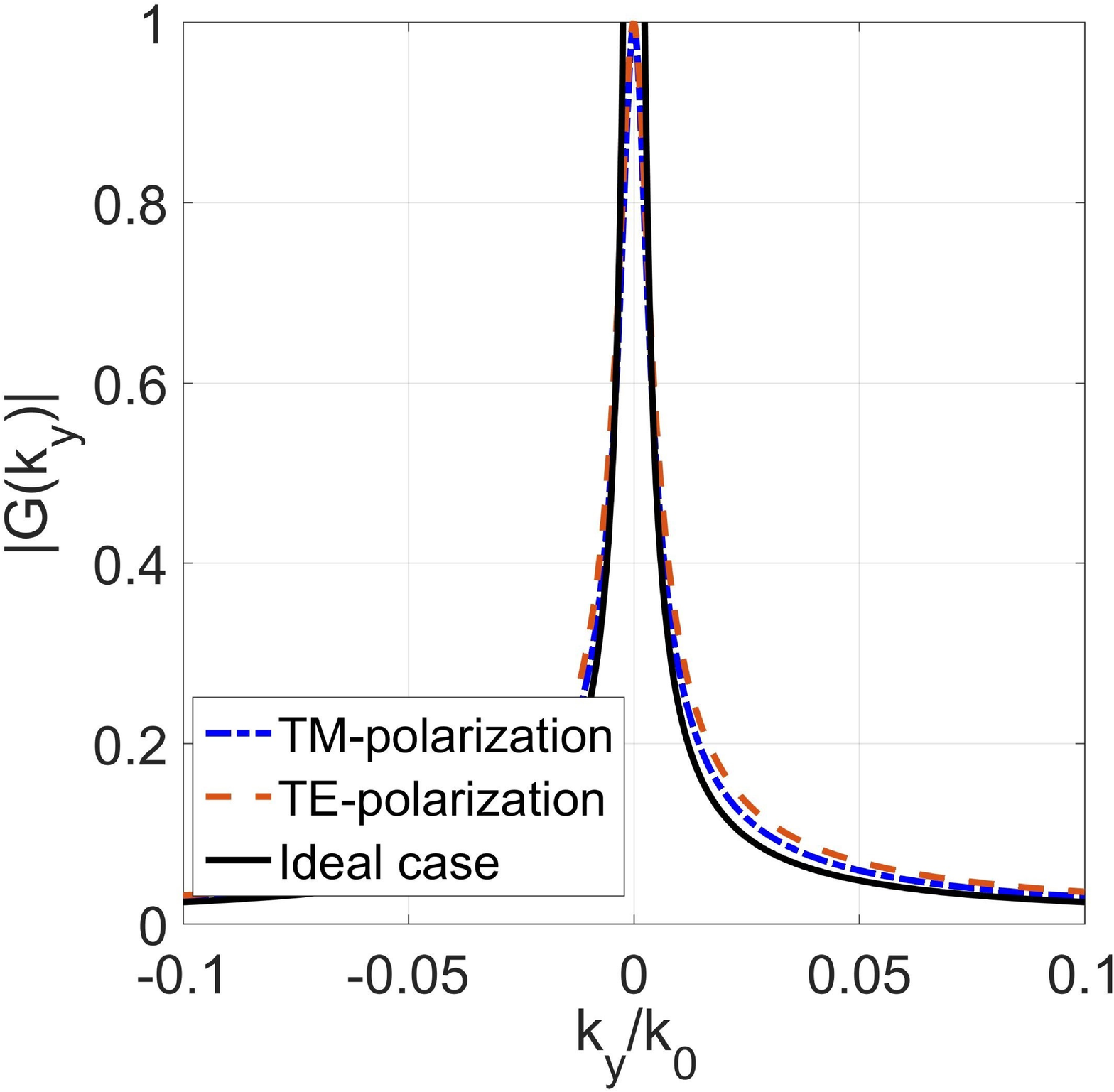}}%
	\subfigure[]{%
		\label{fig:224}%
		\includegraphics[height=2.5in]{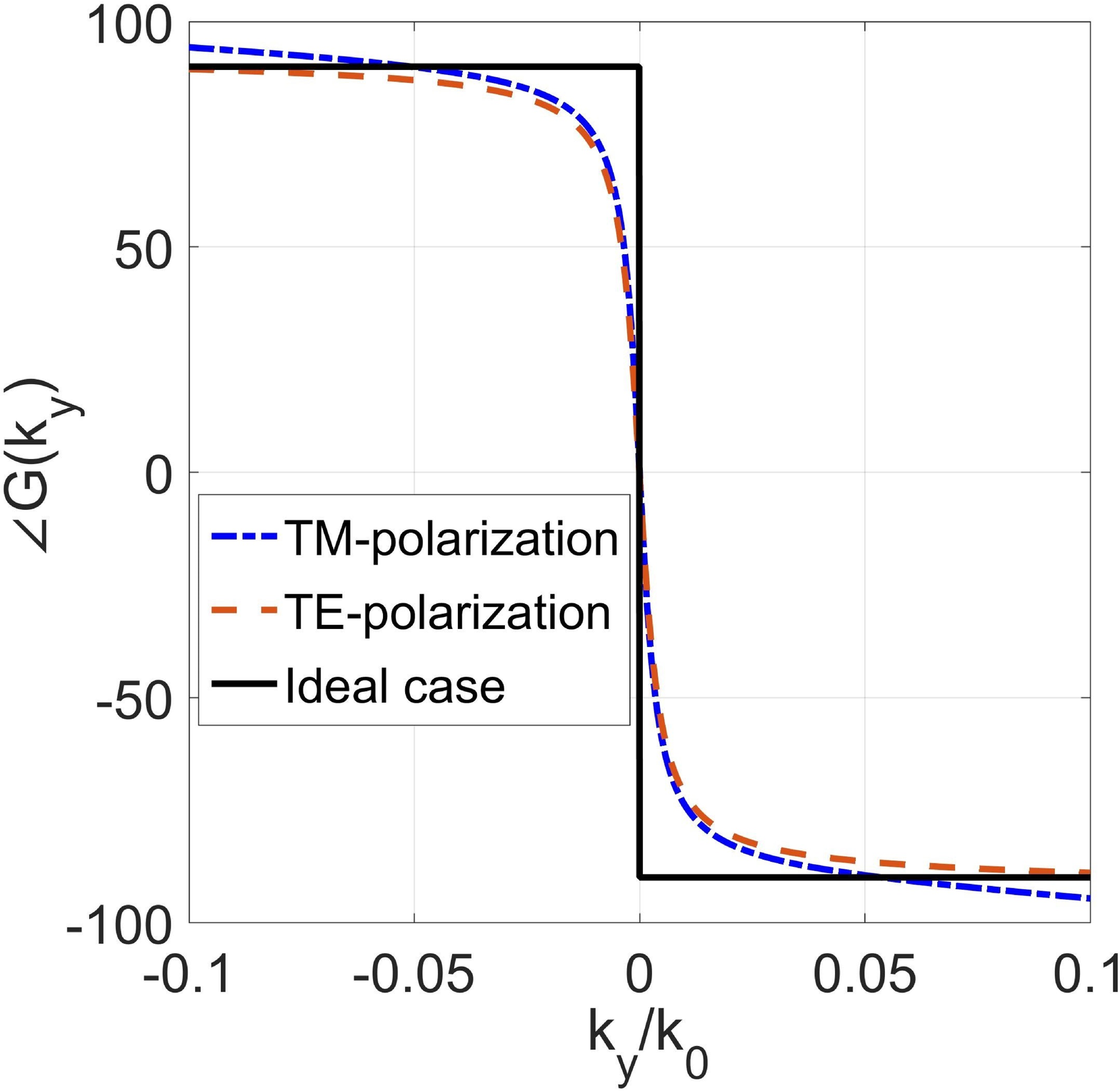}}%
	\caption{ The amplitude and phase of the TE/TM Green's function associated with the polarization-insensitive single-operator computer for first-order (a), (b) derivative and (c), (d) integration operators.   }
	\label{fig:4}
\end{figure*}

\subsection{Polarization multiplexed mathematical operations }
Here, as illustrated in Fig. 4, we propose the concept of polarization multiplexed analog computing whereby the metasurface computer features adjustable operator, which is determined by the polarization state of the incident field. In this way, the aforementioned first-order derivation and integration conditions may be individually satisfied for one of the TE and TM polarizations. As the metasurface computer can be potentially anisotropic, it may response differently for two orthogonal polarizations (TE and TM) while the incident wave polarization selects the mathematical operator. The related configuration is depicted in Fig. 4a. As an example, let us consider the following synthesis problem: find the susceptibilities of a polarization-multiplexed metasurface computer whereby with the TE polarization oblique incidence ($\theta_0=45$), the observed transmission field is formed by the first-order integration of the incident beam profile and when the oblique incident wave has TM polarization, the reflection window is nothing but the first-order derivative of the incident beam profile. This leads to a system of two equations in six unknown susceptibilities, which can be easily solved. At this 
stage, the metasurface computer is synthesized, with the closed-form susceptibilities that are not shown here for the sake of conciseness. Shining a right or left circular polarization (RCP or LCP) wave obliquely onto the metasurface excites both functionalities of the metasurface, as it has both the TE and TM field components (see Fig. 4a). To demonstrate the flexibility of the metasurface computer in executing polarization-multiplexed spatial computing, a circularly-polarized Gaussian-shaped electric field with a spatial bandwidth of $W=0.2k_0$ and incident angle ($\theta_0=45$) is considered as the input field. The reflected/transmitted beams of different polarizations are plotted in Fig. 4b. The results confirm the multiplexing capability of the metasurface by switching the polarization status of incident waves where upon illumination by a circularly polarized oblique incidence, the designed metasurface has the ability to produce both first-order derivative and integration of the incident beam profiles at reflection and transmission modes, respectively. In this section, we have presented a method allowing for the imposition of two independent mathematical operation on any pair of orthogonal states of polarization so that the operator of metasurface can change depending on the polarization status of the incident field. This paper, to the best of our knowledge, is the first work addressing such a multi-functionality that significantly expands the scope of metasurface computers and allowing for new polarization switchable analog computing schemes. In this way, different mathematically processed scattered beam profiles can be obtained by just switching the polarization of the incident wave.  
	\begin{figure*}%
	\centering
	\subfigure[]{%
		\label{fig:31}%
		\includegraphics[height=2.3in]{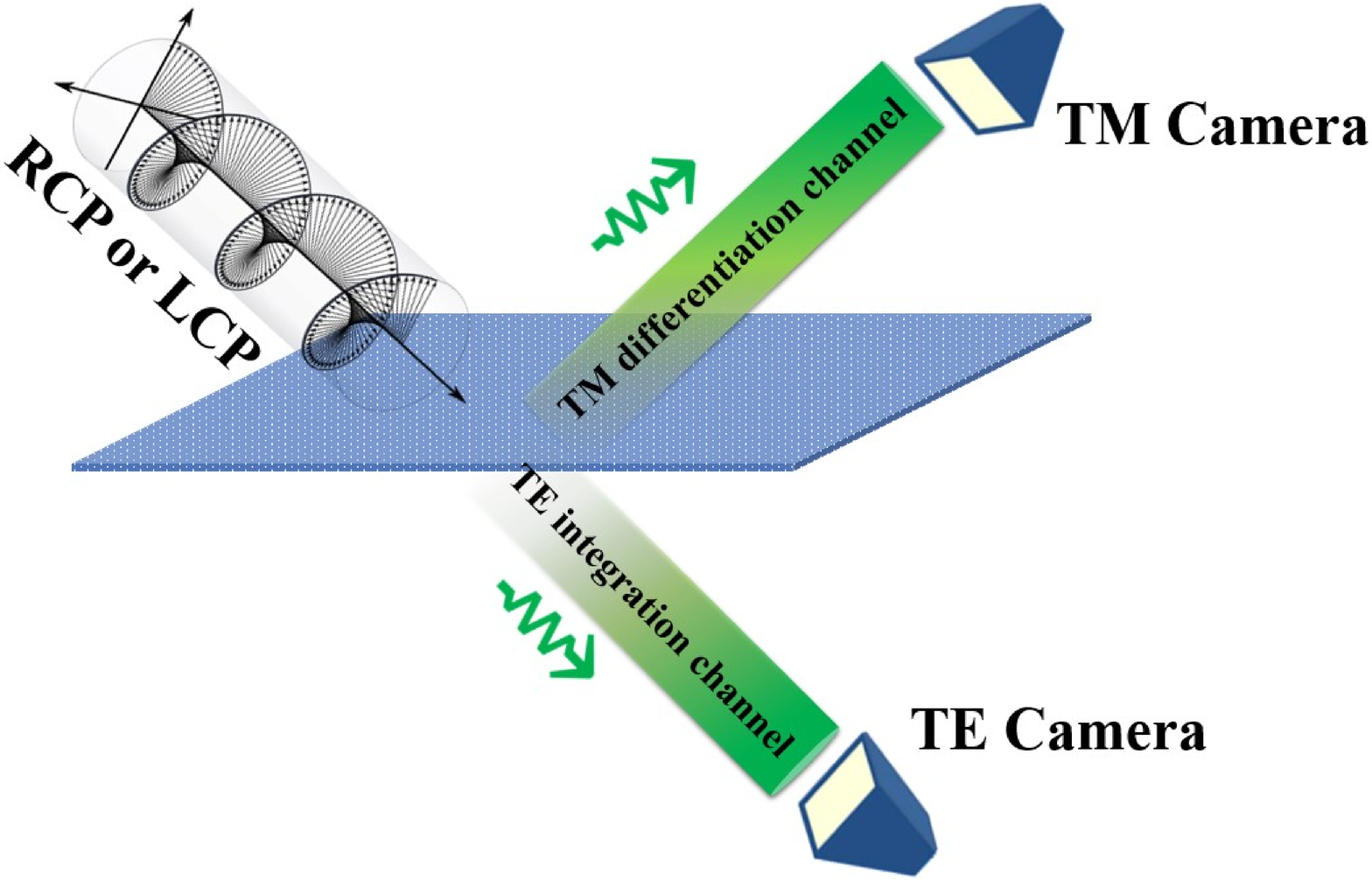}}%
	\subfigure[][]{%
		\label{fig:32}%
		\includegraphics[height=2.3in]{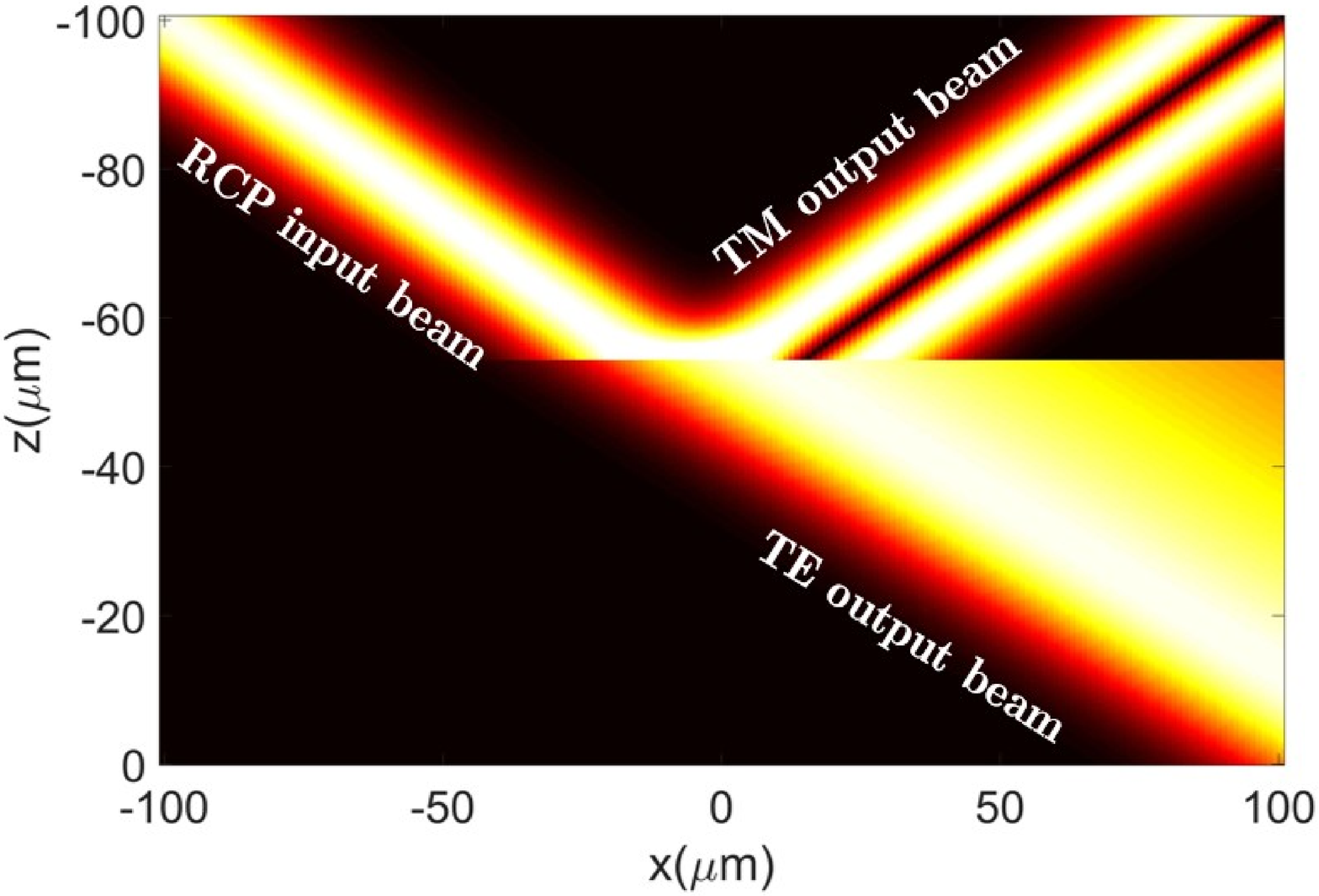}}%
	\qquad
	\caption{(a) The polarization-multiplexed metasurface computer under the illumination of a circularly polarized incident beam where (b) the integration of the TE-polarized part of the input field is fully transmitted while the derivative of the TM-polarized part of the input field is fully reflected by the metasurface computer. }
	\label{fig:4}
\end{figure*}
\subsection{Multi-operator angle multiplexed spatial analog computation }
In this section, we explore a multi-operator metasurface computer of Fig. 5a, which enables multiple and independent parallel optical signal processing when illuminated from different directions, the other great capability which is not previously reported in the literature. The realization of both differentiator and integrator within a single metasurface relies upon modulating the surface susceptibility tensors so that the Green's functions associated with reflection and transmission of metasurface emulates the corresponding operators of choice at different directions. In this way, the integration and derivation conditions must be separately but simultaneously occur for different incident wave angles. The synthesis of a metasurface performing such angle-selective multiple operations requires a number of degrees of freedom, which are granted by leveraging bianisotropy, and using normal susceptibilities. The legitimate question of whether normal polarizations are useful in bringing new functionalities to metasurfaces, or could they be simply ignored and replaced by purely tangential polarizations, was then raised \cite{achouri2018influence,achouri2015general}. As it was shown \cite{achouri2018influence}, normal polarizations may indeed by ignored if a metasurface is always excited with the same illumination conditions, e.g. same incidence angle. However, if the incidence angle is changed, like here, then the presence of normal polarizations do play a role in the scattering behavior of metasurfaces and should not be ignored. The experimental characterization of metasurfaces with normal polarizability has been previously demonstrated in the literature \cite{zaluvski2016analytical} and it is within the realm of current fabrication technologies. To do so, we now take the off-diagonal components of the susceptibility tensor into account and continue the synthesis in such a way that the operator of metasurface is a function of the incident polarization. The only susceptibilities that are relevant to the problem of Fig. 5a for TM polarization are 

\begin{align}
	& \overline{\overline{{{\chi }}}}_{\text{ee}}=\left( \begin{matrix}
		\begin{matrix}
			\begin{matrix}
				\chi _{\text{ee}}^{xx}  \\
				0  \\
				\chi _{\text{ee}}^{zx}  \\
			\end{matrix} & \begin{matrix}
				0  \\
				\chi _{\text{ee}}^{yy}  \\
				0  \\
			\end{matrix}  \\
		\end{matrix} & \begin{matrix}
			\chi _{\text{ee}}^{xz}  \\
			0  \\
			\chi _{\text{ee}}^{zz}  \\
		\end{matrix}  \\
	\end{matrix} \right),\text{    }\overline{\overline{{{\chi }}}}_{\text{em}}=\left( \begin{matrix}
		\begin{matrix}
			\begin{matrix}
				0  \\
				0  \\
				0  \\
			\end{matrix} & \begin{matrix}
				\chi _{\text{em}}^{xy}  \\
				0  \\
				\chi _{\text{em}}^{zy}  \\
			\end{matrix}  \\
		\end{matrix} & \begin{matrix}
			0  \\
			0  \\
			0  \\
		\end{matrix}  \\
	\end{matrix} \right) \\ 
	& \overline{\overline{{{\chi }}}}_{\text{me}}=\left( \begin{matrix}
		\begin{matrix}
			\begin{matrix}
				0  \\
				\chi _{\text{me}}^{yx}  \\
				0  \\
			\end{matrix} & \begin{matrix}
				0  \\
				0  \\
				0  \\
			\end{matrix}  \\
		\end{matrix} & \begin{matrix}
			0  \\
			\chi _{\text{me}}^{yz}  \\
			0  \\
		\end{matrix}  \\
	\end{matrix} \right)\text{,   }\,\,\text{  }\overline{\overline{{{\chi }}}}_{\text{mm}}=\left( \begin{matrix}
		\begin{matrix}
			\begin{matrix}
				0  \\
				0  \\
				0  \\
			\end{matrix} & \begin{matrix}
				0  \\
				\chi _{\text{mm}}^{yy}  \\\nonumber
				0  \\
			\end{matrix}  \\
		\end{matrix} & \begin{matrix}
			0  \\
			0  \\
			0  \\
		\end{matrix}  \\
	\end{matrix} \right) \\\nonumber
\end{align}

The components of the susceptibility tensors are intentionally selected so that the synthesized metasurface is not able to rotate the polarization of the incident wave (polarization-preserved specification). For such a metasurface, the relations between the components of tonsorial Green's functions and susceptibilities are given in
Eqs. (25-27). The other Green's functions pertaining to TM polarization are not given here for the sake of brevity.  From Eqs. (25-27), one can conclude that combining the effects of $\chi _{\text{ee}}^{zy}$ and $\chi _{\text{ee}}^{yz}$, and $\chi _{\text{me}}^{yx}$ and $\chi _{\text{em}}^{xy}$ yields a metasurface which can be potentially asymmetric in both reflection and transmission responses. The presence of $\chi _{\text{ee}}^{zz}$ does not affect the asymmetry as this susceptibility is related to $k_{y}^{2}$ which is a symmetric function of $\theta$ \cite{achouri2018influence}. The presence of  $\chi _{\text{em}}^{zy}$ and $\chi _{\text{em}}^{yz}$ plays a similar role to that of $\chi _{\text{ee}}^{xz}$ and $\chi _{\text{ee}}^{zx}$  since both sets of susceptibilities are related to the presence of ${{k}_{y}}$. As the proof of concept, we present two illustrative examples of angle-multiplexed spatial analog computing. First, a multichannel metasurface-computer was designed to fully reflect the 1st-order derivative and the 1st-order integration of the Gaussian input waves of $20^\circ$ and $45^\circ$ illumination angles, respectively. The angular behavior of the associated Green's functions is depicted in Fig. 5b for the illumination angles between $0^\circ$ and $90^\circ$. 
\begin{figure*}
	\begin{widetext}
		\begin{align}
			& \Omega =2[k_{z}^{2}\chi _{\text{ee}}^{yy}+k_{y}^{2}\chi _{\text{ee}}^{zz}-k_0{{k}_{y}}(\chi _{\text{em}}^{zx}+\chi _{\text{me}}^{xz})+{{k_0}^{2}}\chi _{\text{mm}}^{xx}]+{{k_0}^{2}}(\chi _{ee}^{yy}\chi _{\text{mm}}^{xx}-\chi _{\text{em}}^{yx}\chi _{\text{me}}^{xy}) \\ \nonumber
			& -j{{k}_{z}}[k_{y}^{2}(\chi _{\text{ee}}^{yz}\chi _{\text{ee}}^{zy}-\chi _{\text{ee}}^{yy}\chi _{\text{ee}}^{zz})+4-k_0{{k}_{y}}(\chi _{\text{ee}}^{zy}\chi _{\text{em}}^{yx}+\chi _{\text{ee}}^{yz}\chi _{\text{me}}^{xy}-\chi _{\text{ee}}^{yy}(\chi _{\text{em}}^{zx}+\chi _{\text{me}}^{xz}))] \\ 
			& \tilde{G}_\text{{ref}}^{\text{TE}}({{k}_{y}})=\frac{2}{\Omega }\{k_{y}^{2}\chi _{\text{ee}}^{zz}-k_{z}^{2}\chi _{\text{ee}}^{yy}-{{k}_{z}}[{{k}_{y}}(\chi _{\text{ee}}^{yz}-\chi _{\text{ee}}^{zy})-k_0(\chi _{\text{em}}^{yx}-\chi _{\text{me}}^{xy})]-k_0{{k}_{y}}(\chi _{\text{em}}^{zx}-\chi _{\text{me}}^{xz})+{{k_0}^{2}}\chi _{\text{mm}}^{yy}\} \\  
			& \tilde{G}_\text{{tran}}^{\text{TE}}({{k}_{y}})=\frac{j{{k}_{z}}}{\Omega }\{k_{y}^{2}(\chi _{\text{ee}}^{yz}\chi _{\text{ee}}^{zy}-\chi _{\text{ee}}^{yy}\chi _{\text{ee}}^{zz})+(2j-k_0\chi _{\text{em}}^{yx})(2j-k_0\chi _{\text{me}}^{xy}) \\ \nonumber
			& +{{k}_{y}}[\chi _{\text{ee}}^{zy}(2j-k\chi _{\text{em}}^{yx})+\chi _{\text{ee}}^{yz}(2j-k\chi _{\text{me}}^{xy})+k\chi _{\text{ee}}^{yy}(\chi _{\text{em}}^{zx}-\chi _{\text{me}}^{xz})]-{{k_0}^{2}}\chi _{\text{ee}}^{yy}\chi _{\text{mm}}^{xx}\} \\\nonumber 
		\end{align}
	\end{widetext}
\end{figure*}
\begin{figure*}%
	\centering
	\subfigure[]{%
		\label{fig:41}%
		\includegraphics[height=2in]{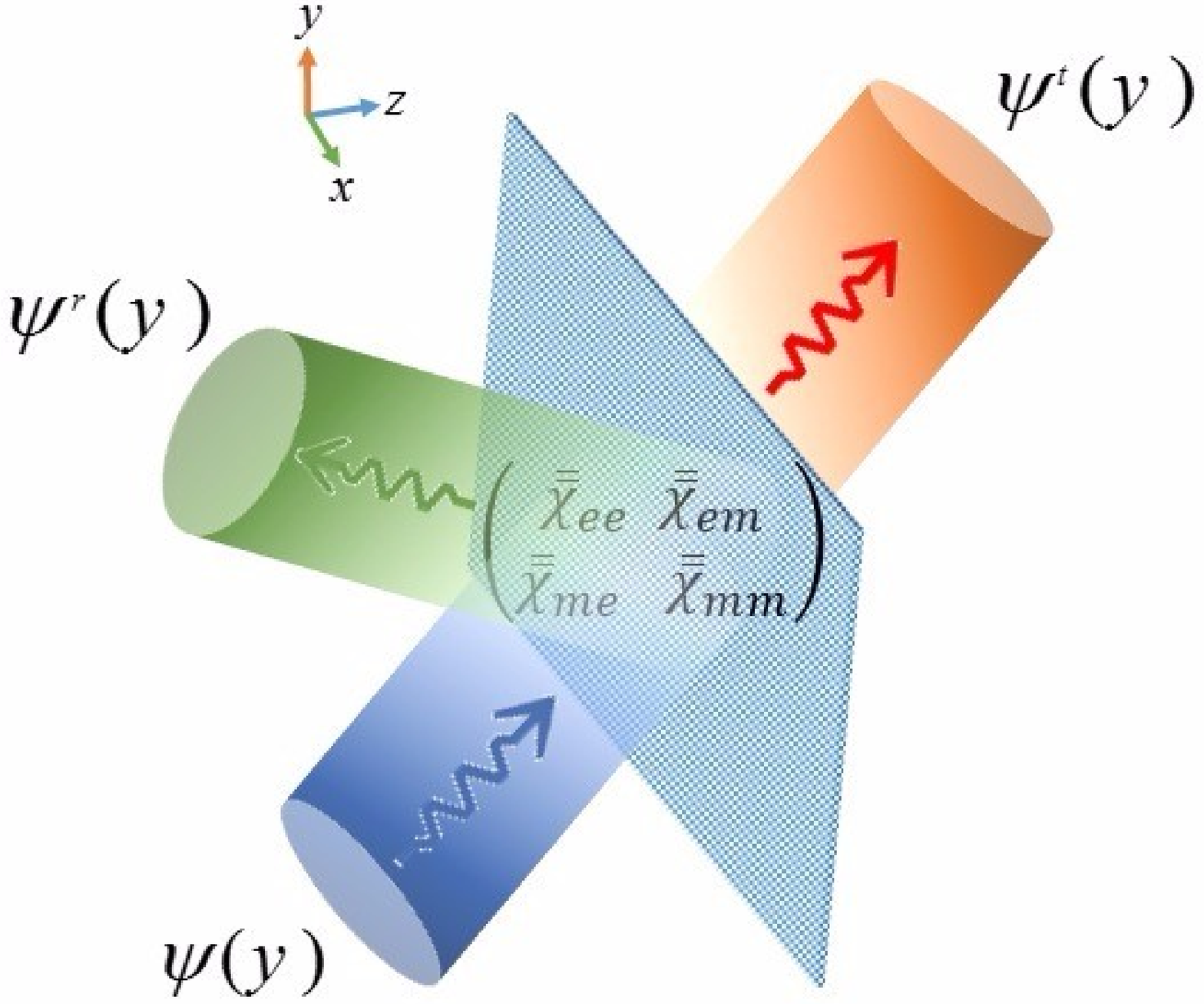}}%
	\subfigure[][]{%
		\label{fig:42}%
		\includegraphics[height=2in]{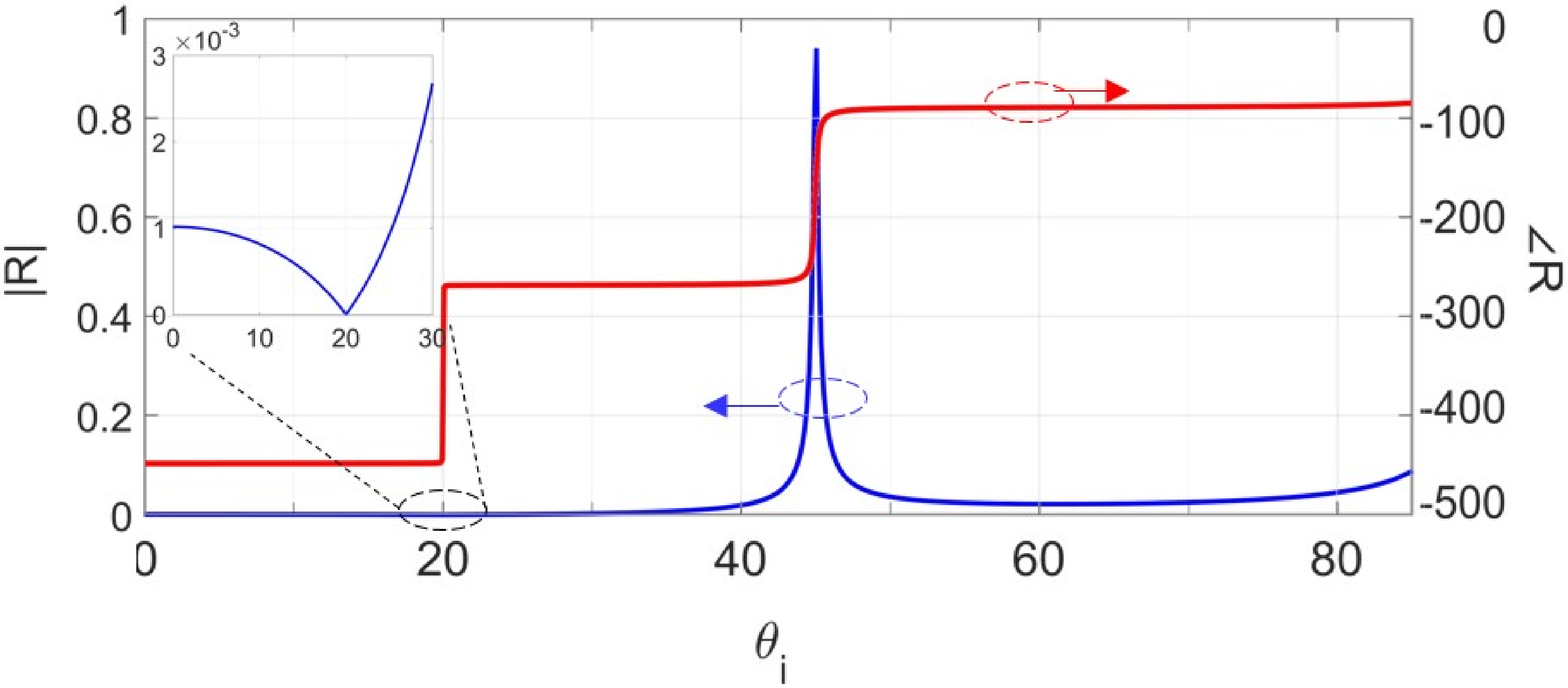}}%
	\qquad
	\caption{(a) The schematic view of an angle-multiplexed analog computing scheme realized by metasurface susceptibility tensors. (b) The reflectivity of the designed metasurface upon illumination by different incident wave angles.  }
	\label{fig:4}
\end{figure*}From this figure, one can clearly see a first-order zero and a first-order peak appearing at around $20^\circ$ and $45^\circ$ of the reflection diagram, respectively, implying that the 1st-order derivative and integration of the Gaussian incident profile will be noticed at reflection around $20^\circ$ and $45^\circ$, respectively. For comparison, Figs. 6a,b display the output beams reflected by the metasurface computer upon different incident wave angles. The results confirm the appropriate performance of the metasurface in performing different mathematical operations on the input signals coming from different directions, thereby creating multiple independent computing channels. As another example, an angle-multiplexed metasurface computer projecting the 1st-order integration and the 1st-order derivative of the input Gaussian signal under $20^\circ$ and  $45^\circ$ illumination angles, respectively was investigated and characterized in the transmission mode. For the sake of briefness, we avoid giving the associated Green's functions, but the transmitted fields upon illumination by different oblique incidences of $20^\circ$ and  $45^\circ$ are plotted in Fig. 6c, d. As can be observed, similar to the previous case, the results of the output beams excellently corroborate the accurate performance of the multi-operator metasurface 
computer. As an important consequence, any two different analog computation masks can be embedded in a single multi-operator metasurface that can be independently accessed upon different pre-determined illumination angles. The homogeneous design drastically simplifies the metasurface implementation since only one unit cell has to be realized and repeated periodically. From a technological point of view, this is a novel class of analog computing platforms, opening the path towards ultra-compact multifunctional flat computers that are not feasible otherwise.
	\begin{figure*}%
	\centering
	\subfigure[]{%
		\label{fig:411}%
		\includegraphics[height=2.1in]{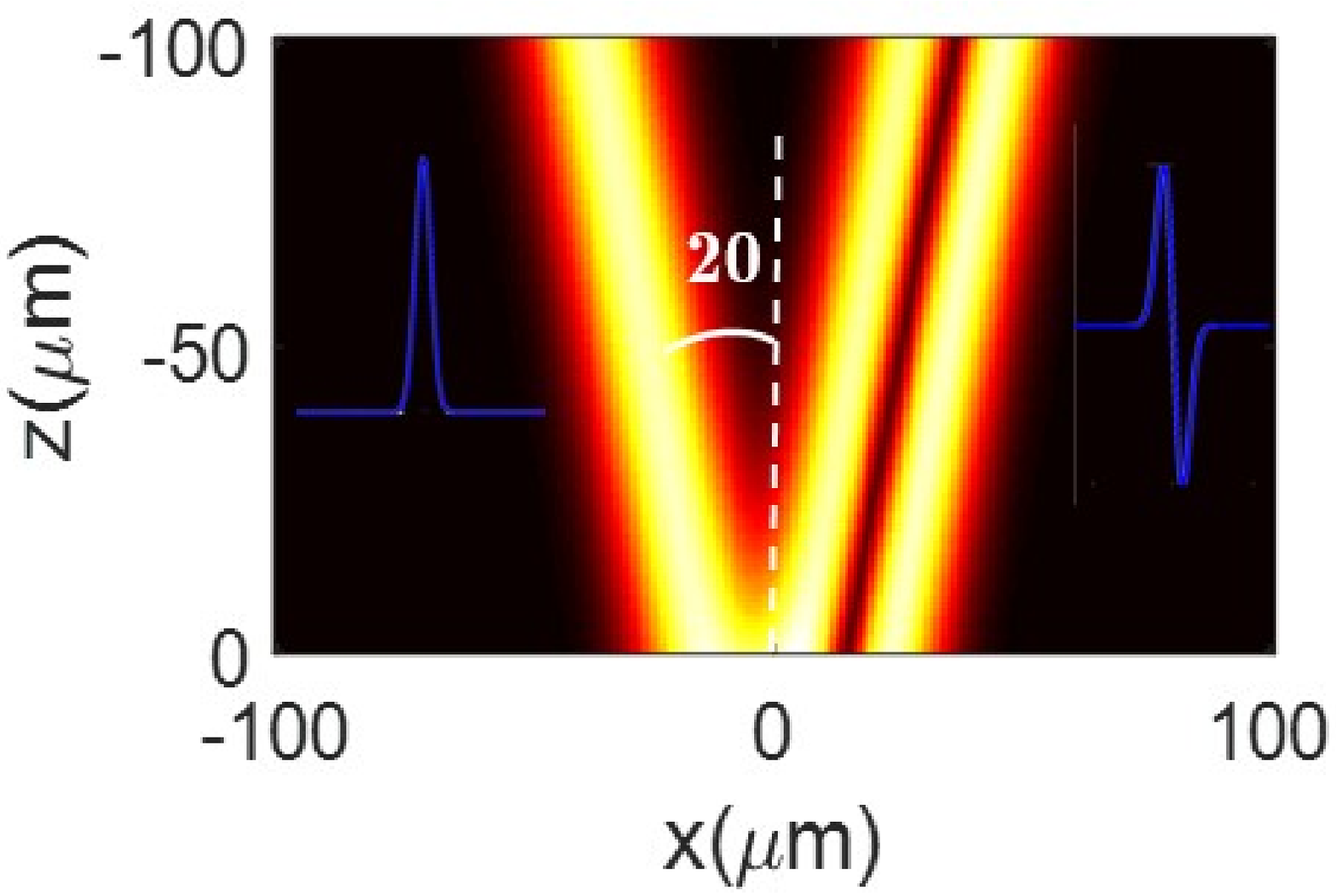}}%
	\subfigure[]{%
		\label{fig:422}%
		\includegraphics[height=2.15in]{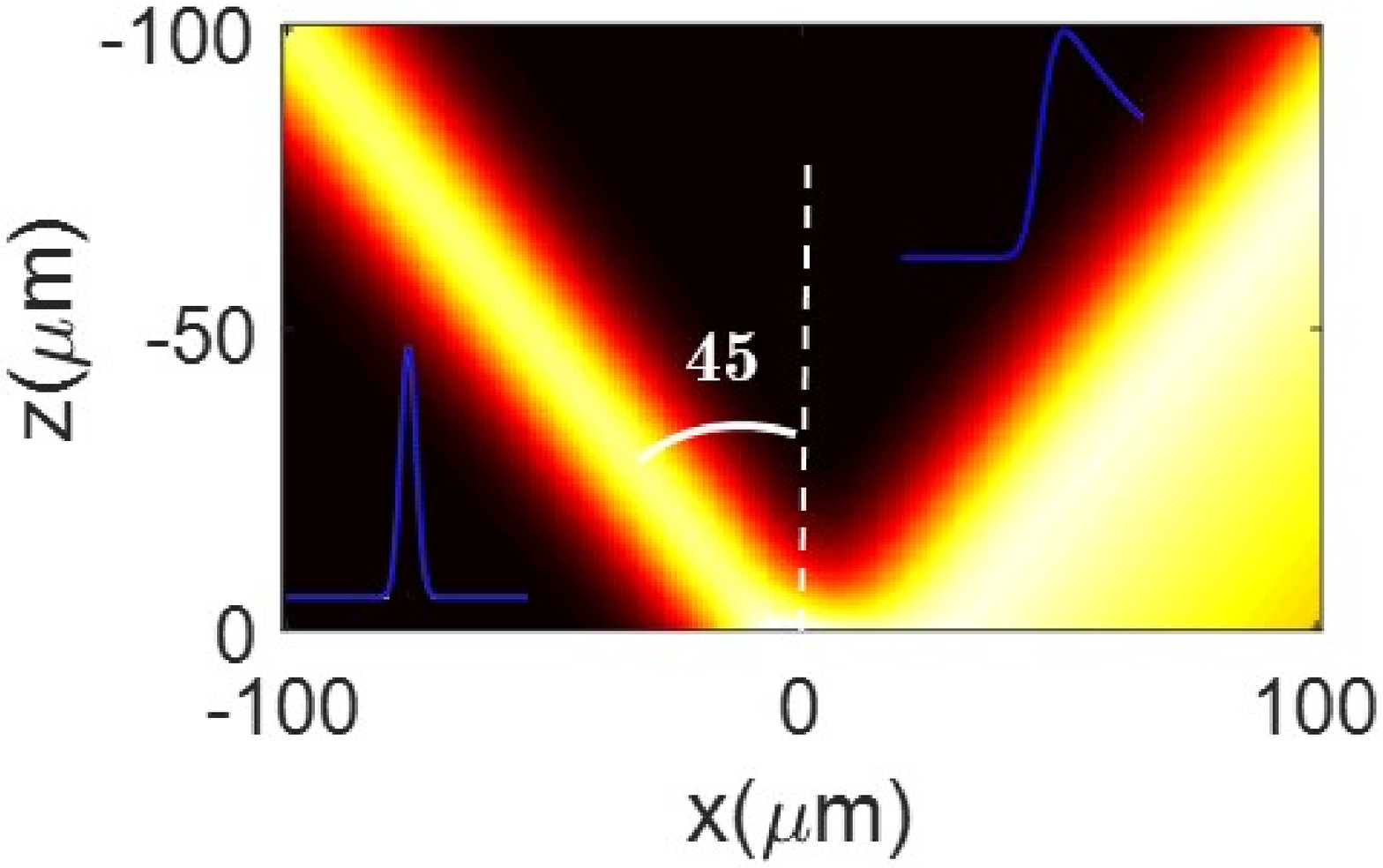}}%
	\qquad
	\subfigure[]{%
		\label{fig:433}%
		\includegraphics[height=2.1in]{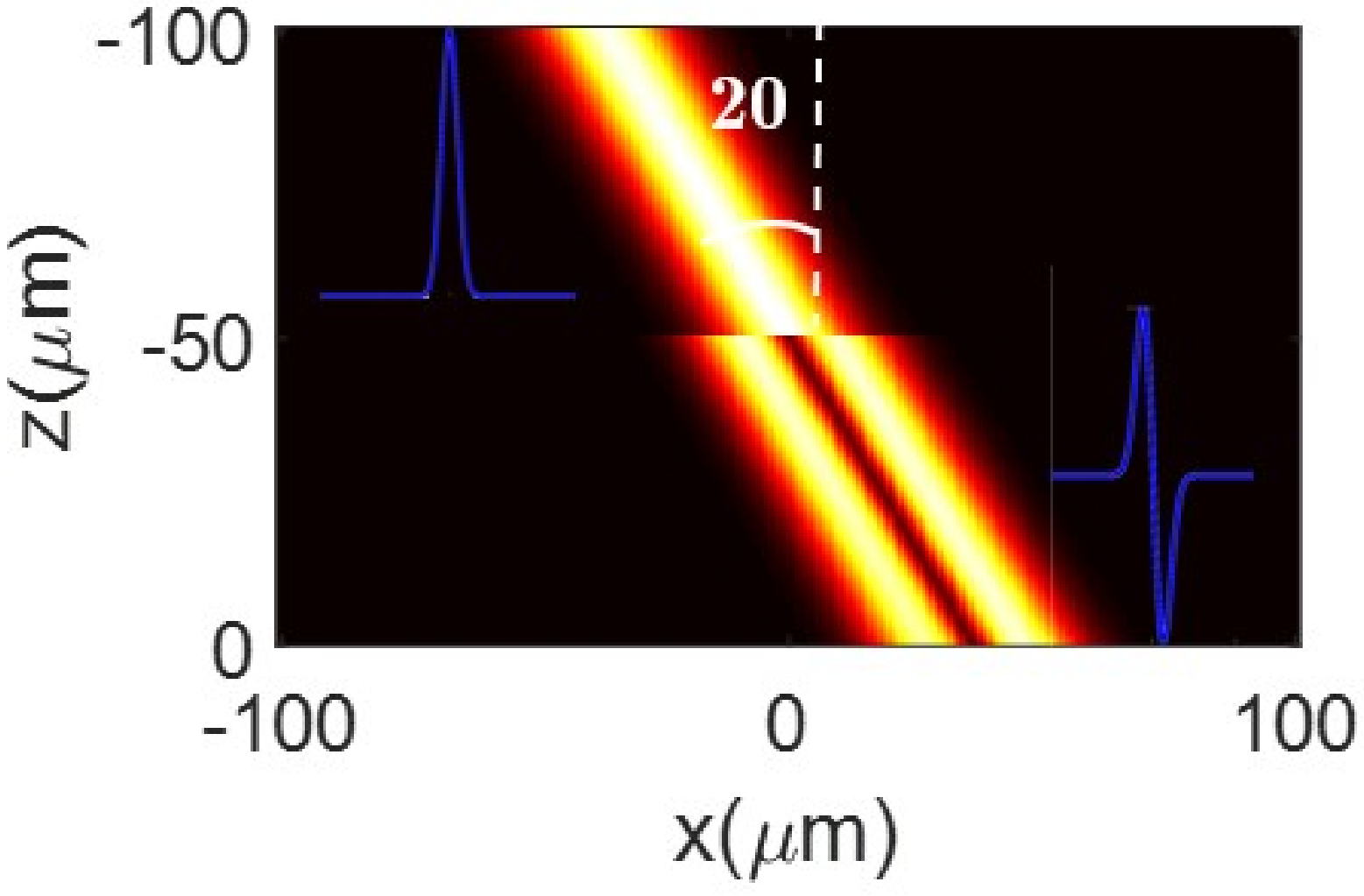}}%
	\subfigure[]{%
		\label{fig:444}%
		\includegraphics[height=2.05in]{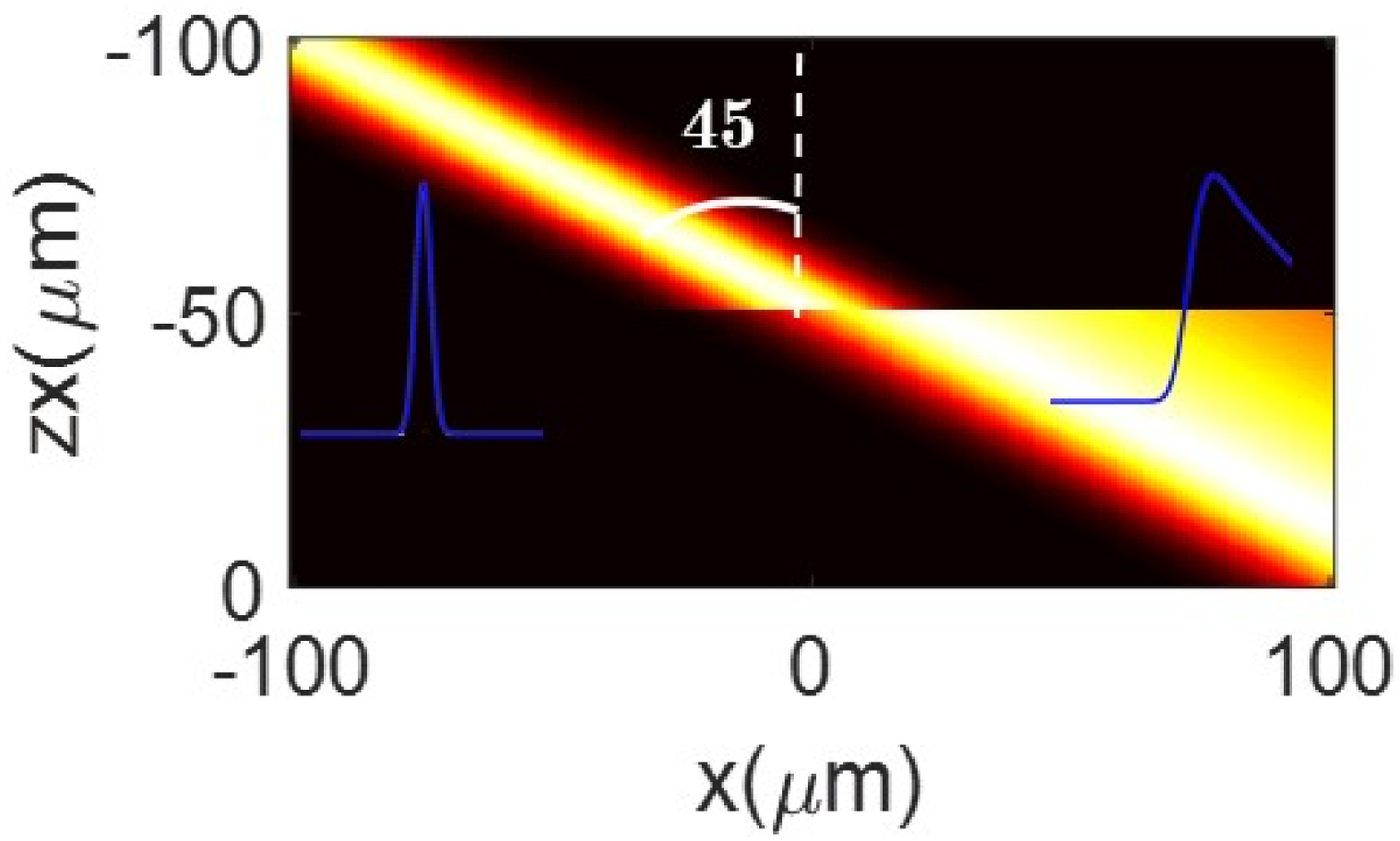}}%
	\caption{  In the first example, the designed metasurface fully reflects the (a) derivative and (b) integration of the input field upon illumination by an oblique incidence of $20^\circ$ and $45^\circ$, respectively. In the second angle-multiplexed design, the metasurface spatial computer fully transmits the (c) derivative and (d) integration of the input field upon illumination by an oblique incidence of $20^\circ$ and $45^\circ$, respectively. }
	\label{fig:4}
\end{figure*}
\subsection{Reflection symmetry-broken mathematical operations }
Up to now, all the associated Green's functions have an even symmetry around the normal incidence (the reflectivity of the conventional structures are even functions of $k_y$). Hence, as a great restriction upon illumination by a normal incidence, the conventional analog computers can only execute those spatial functionalities whose mathematical operator is an even function of ${{k}_{y}}$ . However, there are many appealing and important Green’s functions having odd symmetry in the Fourier domain such as first-order derivative and integration operators. In the previous sections, such an inherent symmetry was broken by a rotated configuration or, in another sense, by oblique incidence. Nevertheless, in some practical scenarios, it may not be possible to have the necessary facilities to excite the metasurface computer with a specified stable obliquely incident wave. In this contribution, our purpose is to propose a simple design to tackle such a drawback and implement the Green’s functions having either odd or even symmetry without resorting to the oblique incidences. In fact, in current designs, regardless of the geometry of the constituent particles, the reciprocity necessarily makes the reflection coefficients to have a ${{\sigma }_{z}}$ symmetry, while the transmission coefficients have a ${{C}_{2}}$ symmetry \cite{achouri2018influence} the same symmetries as those of its scattering particles. In this perspective, we will discuss how relaxing the enforcing reciprocity conditions on the metasurface susceptibilities can break the unwilling odd symmetry of transfer function in reflection mode. 
 	\begin{figure*}%
	\centering
	\subfigure[]{%
		\label{fig:51}%
		\includegraphics[height=2in]{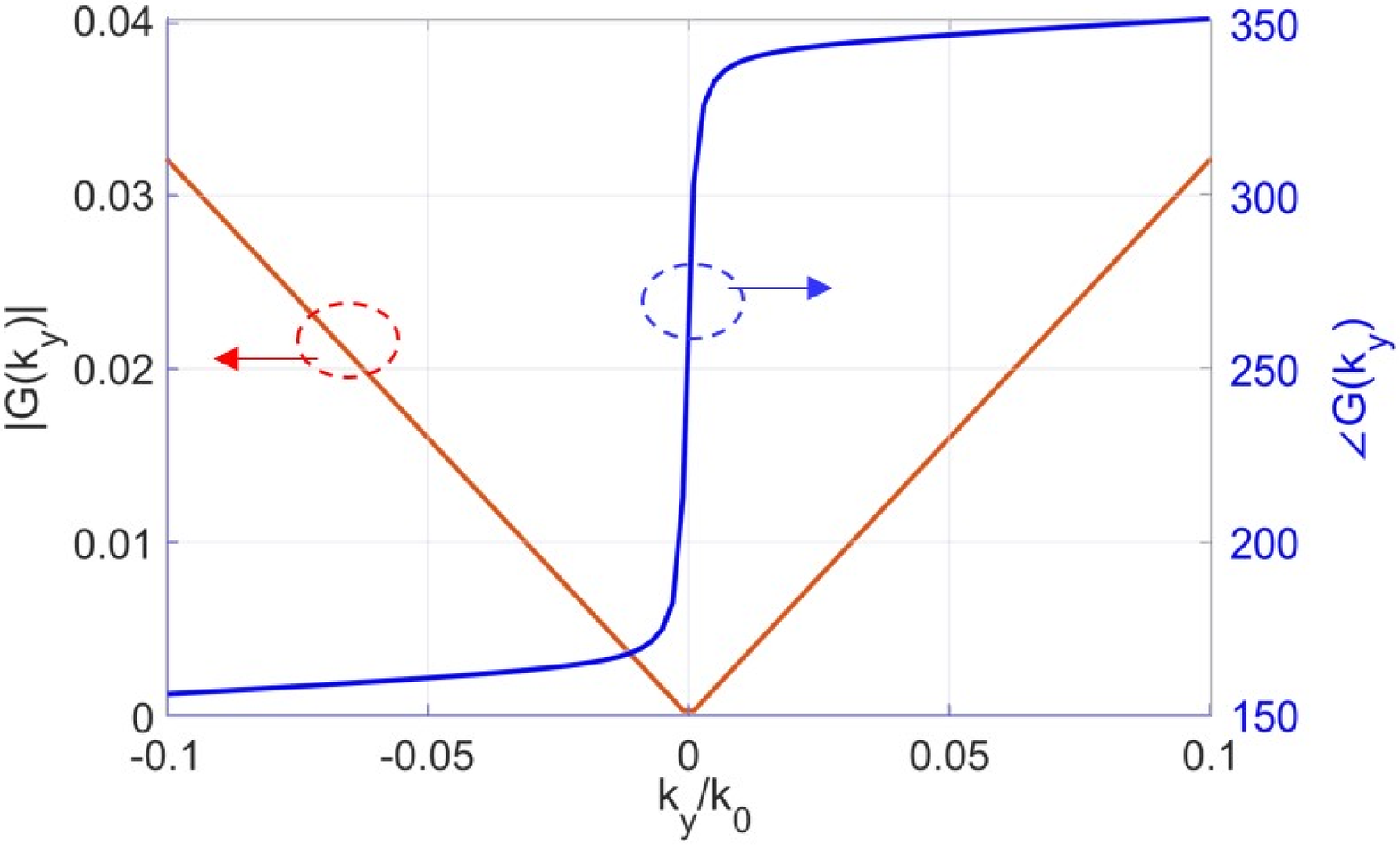}}%
	\subfigure[]{%
		\label{fig:25}%
		\includegraphics[height=2.15in]{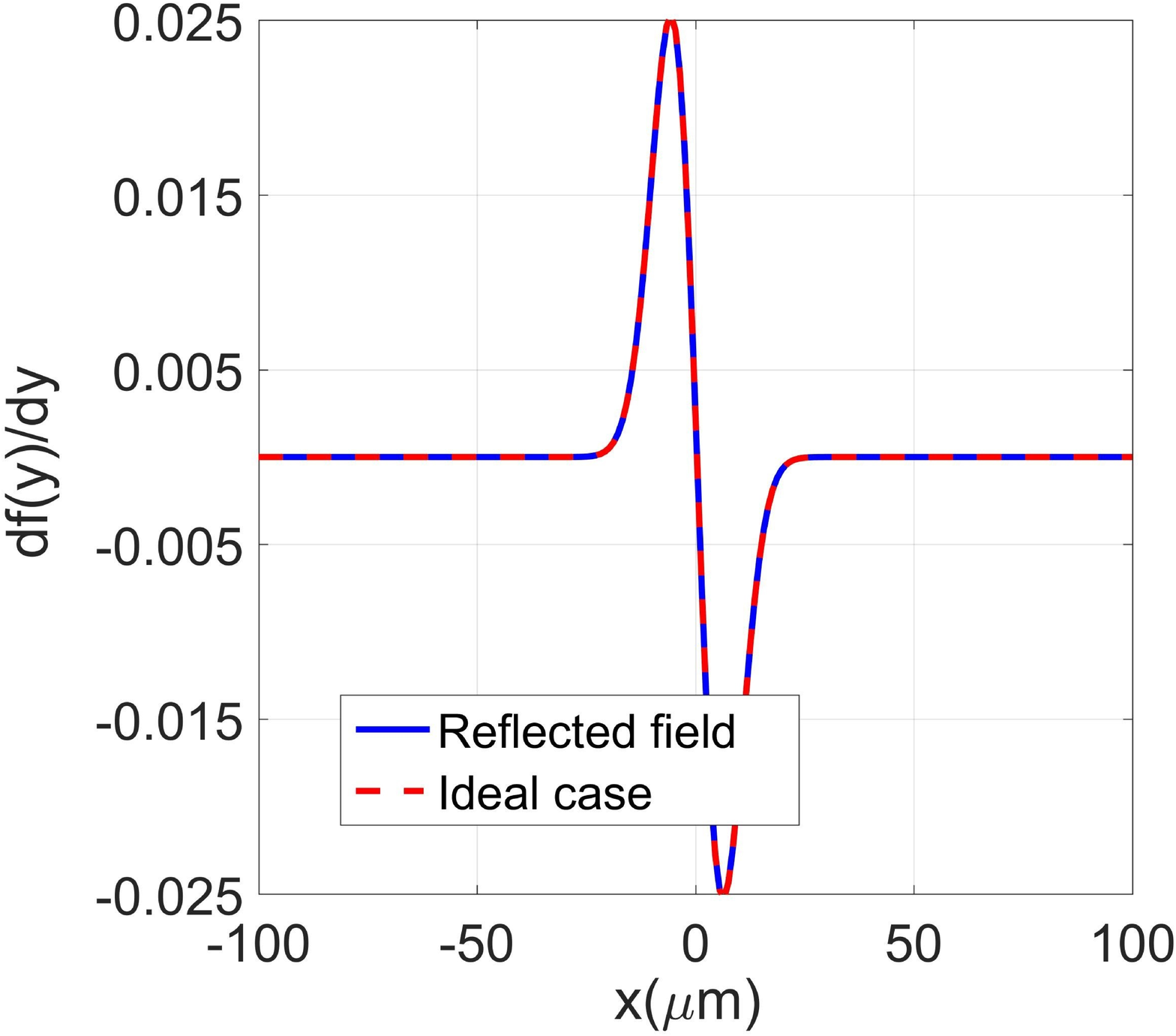}}%
	\qquad
	\subfigure[]{%
		\label{fig:35}%
		\includegraphics[height=1.5in]{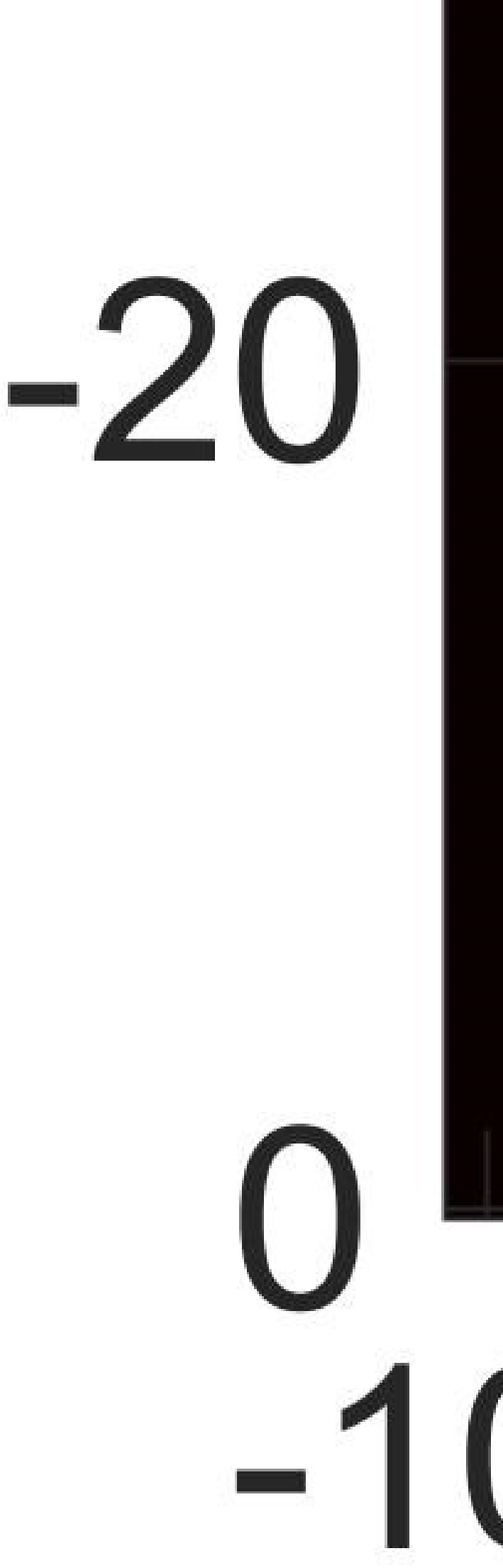}}%
	\subfigure[]{%
		\label{fig:45}%
		\includegraphics[height=1.5in]{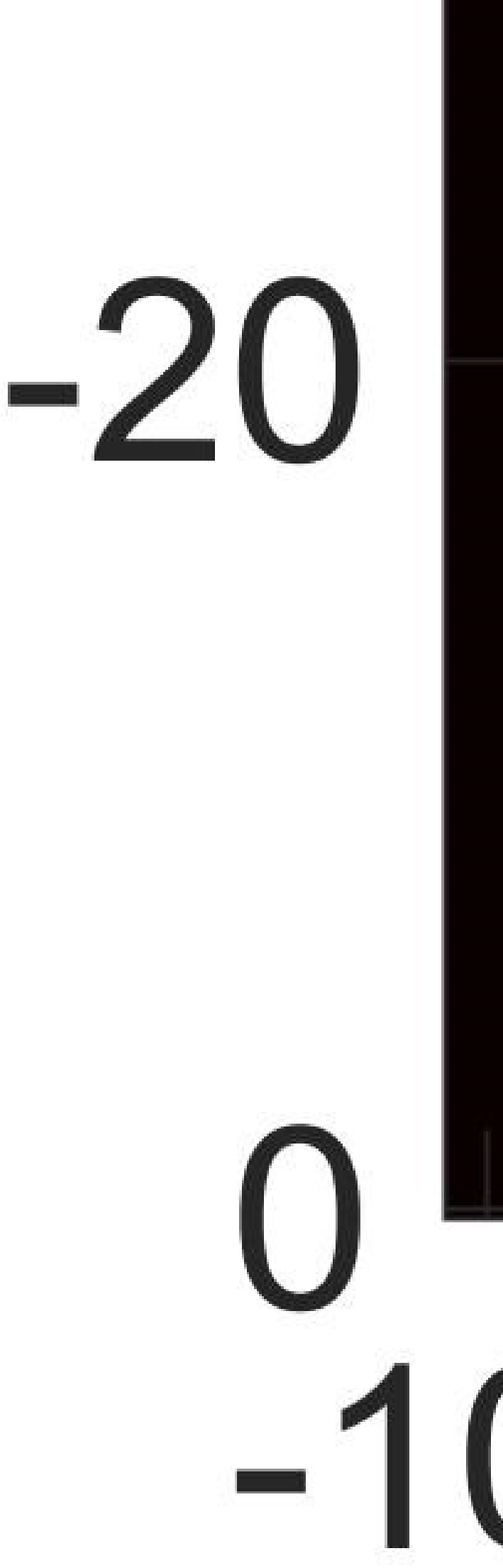}}%
	\caption{ Exploiting the non-reciprocal metasurface computer for breaking the even-reflection symmetry of the operator of choice at normal incidences. (a) The amplitude and phase of the associated Green's function. (b) The reflected field profile of the derivation of the input field upon illumination by a normal incidence. (b) The normally incident field profile. (d) The reflected field profile.  }
	\label{fig:5}
\end{figure*}
In this case, a nonreciprocal metasurface must somehow be able to break the reciprocal angular scattering symmetries \cite{achouri2018influence}. It is important to consider that the asymmetry of this anisotropic metasurface, whether it is reciprocal or not, is directly related to the presence of the term  ${{k}_{y}}$ and not that of $k_{y}^{2}$ since the latter is a symmetric function of $\theta $. Upon inspection of Eq. (25-27), we see that the presence of ${{k}_{y}}$ is related to the presence of $\chi _{\text{ee}}^{yz}$ and $\chi _{\text{ee}}^{zy}$ while $\chi _{\text{ee}}^{zz}$ is essentially related to $k_{y}^{2}$. It follows that the susceptibility components $\chi _{\text{ee}}^{zy}$ and $\chi _{\text{ee}}^{yz}$ are responsible for the asymmetric angular scattering of this metasurface \cite{achouri2018influence}. Recently, Alu. et al explores a reciprocal nonlocal metasurface enabling the optical analog computations for normal incidences but only in the transmission mode \cite{kwon2018nonlocal}. As comprehensively discussed in this paper, it would be impossible to design a reciprocal metasurface providing different reflection coefficients for the angle of incidences $\theta$  and  -$\theta$ (breaking the even reflection symmetry). We would like to note here that non-reciprocity in a broader sense refers to systems that break the Lorentz reciprocity condition \cite{achouri2018influence,lindell1994electromagnetic}. Fundamentally, new surface properties can be realized when these symmetry restrictions are circumvented. Let us now consider the case where the metasurface computer has normal susceptibilities with nonreciprocal feature, i.e. $\chi _{\text{ee}}^{yz}\ne \chi _{\text{ee}}^{zy}$. In this way, the term ${{k}_{y}}$ does not vanish in Eq. (25-27), leading to a nonreciprocal (symmetry-broken) reflection behavior at normal incidence. Let us know look at an illustrative example in which a non-reciprocal metasurface possesses non-reciprocal feature through exhibiting zero-reflection with an odd symmetry around the boresight direction. For the sake of conciseness, the related susceptibility components are not shown here. Fig. 7a plots the corresponding Green's where as expected, the reflection coefficient is asymmetric around broadside (${{k}_{y}}=0$) with a 180 phase change. Let us analyze the performance of both metasurfaces for a Gaussian TM-polarized normal incidence with the normalized spectral beamwidth of $W=0.1k_0$ . The inset of Fig. 7b illustrates the magnitude of the analytically computed derivative, and the magnitude of the reflected beam upon diffraction of Gaussian normally impinges on the non-reciprocal metasurfaces, respectively. Although, due to inherent even-symmetry reflection, the reciprocal metasurface fails to transform the Gaussian incident beam into Hermite-Gaussian modes, the electric field profile reflected by the non-reciprocal metasurface is completely quite identical to the analytically computed first-order derivative within the accuracy of plot as the Pearson's correlation coefficient between them exceeds 0.99. The excellent performance of the metasurface computer upon illumination by normal incidences can be further clarified from the incident and reflected beams shown in Figs. 7c, d. To the best of our knowledge, this paper is the first proposal that reports a zero-thickness surface can perform a reflective odd-symmetry mathematical operation on input signals coming from normal direction. Recent years have witnessed an astonishing progress in design of nonreciprocal devices whose electromagnetic responses alter when the position of source and detector is exchanged. Involving the magneto-optical effect \cite{gusynin2006magneto,minovich2015functional}, nonlinearity \cite{alam2016large,li2017nonlinear,yang2015nonlinear,li2017nonlinear}, space-time modulation \cite{taravati2017nonreciprocal,shaltout2015time,taravati2017nonreciprocal,sounas2017non}, Kerr nonlinearity \cite{yang2015nonlinear,krasnok2017nonlinear,shcherbakov2015ultrafast,esembeson2008high}, or surface-circuit-surface architecture \cite{taravati2017nonreciprocal} can break the inherent time reversal symmetry and yield nonreciprocal surface properties. Besides, metasurfaces supporting out-of-plane surface currents can be represented by spatially-dispersive surface parameters effectively biased through the transverse momentum of the incident wavefront, imitating nonreciprocal phenomena. Therefore, the recent achievements can promise a feasible platform for realizing the proposed metasurface computer executing both odd- and even-symmetry mathematical operations upon illumination by normal incidences.
 \begin{figure}[t]
 	\includegraphics[height=3.05in]{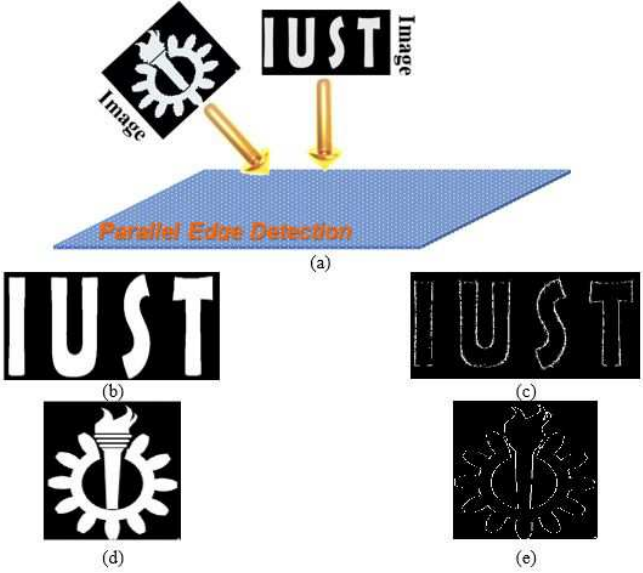}
 	\caption{\label{fig:epsart} (a) Demonstration of the multi-channel parallel edge detection scheme (b) IUST" character and (d) Iran University of Science and Technology logo as the input beams which simultaneously excite the metasurface computer from different directions. (c), (e) Edges of the input images resolved in different angles of transmission and reflection. }
 \end{figure}
\section{Parallel edge detection scheme }

As an initial step in recognition of objects in image processing, spatial differentiation enables the extraction of boundaries of two regions with different texture characteristics, referred to edge detection. As can be seen from Fig. 8a, we now demonstrate that the metasurface computer designed based on the GCTS-susceptibility approach can be readily exploited as a flexible 1D edge detector. Since it provides the ability to simultaneously perform the spatial differentiation for incident waves of different polarizations and directions, we can use multiple parallel channels for processing various images at the same time. We show different aspects of our differentiator by projecting multiple images on the metasurface with different illumination angles. Although all examples given here are related to 1D configurations, we can simply extend them into 2D scenarios \cite{saba2018two} so that the metasurface computer can perform spatial differentiation along either single or both directions. In the first case, directional selectivity is very beneficial in image processing to determine and classify edge directions \cite{kwon2018nonlocal}. While the latter case reveals the ability to detect edges with different orientations as a significant advantage in comparison to 1D differential operators \cite{pors2014analog}. In our numerical demonstrations, the output images have been computed based on the standard image processing rules \cite{saba2018two}. As the first illustration, to verify the versatility of our analog computer, the metasurface is designed to reflect the 1D derivative of the incident images coming from ${{\theta }_{\text{inc},1}}={{0}^{\circ }}$ and transmit the 1D derivative of the incident image coming from ${{\theta }_{inc,1}}={{60}^{\circ }}$. Furthermore, armed with the designed structure, two different images can be independently involved in edge detection at the same time. Fig. 8b, d depicts the Iran University of Science and Technology logo and the 'IUST' character as two different image fields illuminating the metasurface at the same time for evaluating the edge detection performance. The inside and outside regions of the images have different intensities. Although the intensity of edges is not uniform in different directions, the information about existence of edges of an image in all directions can be provided by considering a threshold amplitude in the output image. The reflected and transmitted images at different directions are shown in Figs. 8c, e, which successfully expose all outlines of the incident images in the horizontal orientations with the same intensity. Eventually, the results provided in this section illustrates that the designed metasurface computer can be served as a multi-channel feasible edge detection platform operating for separate directions and polarizations, which have been not reported, yet. 

\section{conclusion}
In summary, we present an overview of the generalized optical signal processing based on multi-operator metasurfaces synthesized by susceptibility tensors. Through establishing normal susceptibility tensors and/or nonreciprocal components, a polarization- and angle-multiplexed metasurface enabling multiple and independent parallel analog spatial computations when illuminated by differently polarized incident beams from different directions is revolutionary designed. The proposed theoretical framework foretastes that the presented design overcomes the substantial restrictions imposed by previous investigations such as large architectures arising from the need of additional subblocks, slow responses, and most importantly, supporting only the even reflection symmetry operations for normal incidences, working for a certain incident angle or polarization, and executing only single mathematical operation. As one of the hotspots in optical signal processing, different aspects of multichannel edge detection scheme are demonstrated in this paper through simultaneous projecting multiple images on the metasurface of different directions. The numerical results prove that the proposed metasurface computer may be thought of as an efficient and flexible host for being utilized in the field of optical signal processing.


\bibliography{apssamp}

\end{document}